\documentclass[%
 reprint,
superscriptaddress,
 amsmath,amssymb,
 aps,prc
]{revtex4-2}

\usepackage{graphicx}
\usepackage{dcolumn}
\usepackage{bm}
\usepackage{float} 
\usepackage[colorlinks,citecolor=blue]{hyperref}
\usepackage{bookmark}
\graphicspath{{./}{figures/}}

\newcommand{\iso}[2]{\ensuremath{^{#1}\rm{#2}}}
\newcommand{\sun}{\odot}
\newcommand{\msun}{\ensuremath{M_\sun}}
\newcommand{\mev}{\mbox{MeV}}
\newcommand{\kev}{\mbox{keV}}
\newcommand{\second}{\mbox{s}}
\newcommand{\bepa}{\ensuremath{{^{10}\mbox{Be}(p,\alpha)^7\mbox{Li}}}}
\newcommand{\bean}{\ensuremath{{^{10}\mbox{Be}(\alpha,n)^{13}\mbox{C}}}}
\newcommand{\benue}{\ensuremath{{^{10}\mbox{Be}(\nu_e,e^-)^{10}\mbox{B}}}}
\newcommand{\liap}{\ensuremath{{^{7}\mbox{Li}(\alpha,p)^{10}\mbox{Be}}}}

\newcommand{\UTphys}{Department of Physics and Astronomy, University of Tennessee, Knoxville, TN 37996-1200, USA}
\newcommand{\ORNLphys}{Physics Division, Oak Ridge National Laboratory, P.O. Box 2008, Oak Ridge, TN 37831-6354, USA}
\newcommand{\GSI}{GSI Helmholtzzentrum f{\"u}r
 Schwerionenforschung, Planckstra{\ss}e 1, 64291 Darmstadt, Germany}
\newcommand{\TUDa}{Institut f{\"u}r Kernphysik (Theoriezentrum),
    Fachbereich Physik, Technische Universit{\"a}t
    Darmstadt, Schlossgartenstra{\ss}e 2, 64298 Darmstadt, Germany}
\begin{document}

\preprint{PRC}

\title{The role of low-lying resonances for the\texorpdfstring{\bepa}{10Be(p,a)} reaction rate and implications for the formation of the Solar System}

\author{A. Sieverding}
 \email{sieverdinga@ornl.gov}
\affiliation{\ORNLphys}
\author{J. S. Randhawa}%
\affiliation{Department of Physics and The Joint Institute for Nuclear Astrophysics, University of Notre Dame, Notre Dame, Indiana 46556 USA}
\author{D. Zetterberg}
\affiliation{\ORNLphys}
\affiliation{\UTphys}
\author{R. J. deBoer}
\author{T. Ahn}
\affiliation{Department of Physics and The Joint Institute for Nuclear Astrophysics, University of Notre Dame, Notre Dame, Indiana 46556 USA}
\author{R. Mancino}
\affiliation{\TUDa}
\affiliation{\GSI}
\author{G. Martínez-Pinedo}
\affiliation{\GSI}
\affiliation{\TUDa}
\author{W. R. Hix}
\affiliation{\ORNLphys}
\affiliation{\UTphys}
\date{\today}

\begin{abstract}
  Evidence for the presence of short-lived radioactive isotopes when the
  Solar System formed is preserved in meteorites, providing insights
  into the conditions at the birth of our Sun.  A low-mass
  core-collapse supernova had been postulated as a candidate for the
  origin of \iso{10}{Be}, reinforcing the idea that a supernova
  triggered the formation of the Solar System. We present a detailed
  study of the production of \iso{10}{Be} by the $\nu$ process in
  supernovae, which is very sensitive to the reaction rate of the
  major destruction channel, \bepa. With data from recent nuclear
  experiments that show the presence of a resonant state in
  \iso{11}{B} at $\approx$~193~keV, we derive new values for the \bepa\
  reaction rate which are significantly higher than previous
  estimates. We show that, with the new \bepa\ reaction rate, a
low mass CCSN is unlikely to produce enough \iso{10}{Be} to  explain the observed \iso{10}{Be}/\iso{9}{Be} ratio in meteorites, even for a wide range of 
neutrino spectra  considered in our models. These findings point towards spallation reactions induced by 
solar energetic particles in the early Solar System as the origin of \iso{10}{Be}.

\end{abstract}

\maketitle

\section{\label{sec:background}Background}

The presence of now-extinct, short-lived ($T_{1/2}<10\,\rm{Myr}$) radioactive nuclei (SLRs) in the early Solar System (ESS) has been revealed by systematic anomalies in the abundances of their stable daughter isotopes \cite{Lee.Papanastassiou.ea:1977,Reynolds:1960} in meteorites. Since their lifetimes are shorter than the timescales on which the composition of the interstellar medium homogenizes, these radioactive isotopes must have been injected into the Solar System by a nearby nucleosynthesis event or they must have been produced in-situ. Precisely determined abundance ratios of SLRs, thus, provide a very powerful tool to understand the birth environment of our Sun\cite{Dauphas.Chaussidon:2011}.
The requirement to consistently explain all the observed values, however, challenges theoretical models and enrichment scenarios and there is an ongoing debate 
about the sources of individual isotopes (see \cite{Lugaro.Ott.ea:2018} for a recent review).

Recent research has added evidence to the idea that core-collapse supernovae (CCSNe) trigger star formation \cite{Zucker.Goodman.ea:2022,Forbes.Alves.ea:2021,Krause.Burkert.ea:2018}, making such events prime candidates for the origin of SLRs, as already suggested more than 40 years ago by \citet{Cameron.Truran:1977}. 
The isotope \iso{10}{Be} with a half-life of $T_{1/2}= 1.4\,\rm{Myr}$ \cite{Chmeleff:2010}
has mostly been studied in the context of non-thermal, in-situ nucleosynthesis due to cosmic rays or high energy particles from the Sun in its early phases. Such processes have been shown to possibly produce enough \iso{10}{Be} to explain the ESS value \cite{Fukuda.Hiyagon.ea:2019,Tatischeff.Duprat.ea:2014,Duprat.Tatischeff.ea:2007,Desch.Connolly.ea:2004,Lee.Shu.ea:1998}, but rely on assumptions about the cosmic ray spectra, the properties of young stellar objects, and mixing in the ESS.
However, \iso{10}{Be} is also produced by the $\nu$ process in CCSN explosions \cite{Woosley.Hartmann.ea:1990}, i.e., by the interactions of high energy neutrinos emitted from  the nascent proto neutron star with nuclei in the outer shells of the massive star.
The relevant reactions are the neutral-current reaction
\iso{12}{C}$(\nu_x,\nu_x' p p)^{10}$Be and the charged-current reaction \iso{12}{C}$(\bar{\nu}_e,e^+ n p)^{10}$Be. In these reactions, interactions with high-energy neutrinos from the tail of the distribution
populate high-lying states of \iso{12}{C} and \iso{12}{B} that decay by multi-particle emission. These transitions are usually dominated by the giant resonances and the total cross-section is constrained by sum rules \cite{Balasi.Langanke.ea:2015}, making it relatively insensitive to the details of the nuclear structure. The neutrino spectra from the CCSN are, however, a major uncertainty for the production channel.

\citet{Banerjee.Qian.ea:2016} have pointed out that a low-mass CCSN is particularly favorable as source for SLRs in the ESS because the yields of O, Mg, Si, Ca, Fe, and Ni increase steeply with the progenitor mass and would lead to anomalies in the abundances of stable isotopes, which are, however, not observed.
They have further demonstrated
that such a low-mass CCSN of a 11.8~\msun progenitor still
produces a sufficient yield of \iso{10}{Be} to simultaneously explain the \iso{10}{Be}/\iso{9}{Be}, \iso{41}{Ca}/\iso{40}{Ca} and \iso{107}{Pd}/\iso{108}{Pd} isotopic ratios without creating an overabundance of any of the other SLRs for which abundance ratios have been determined (with the exception of \iso{60}{Fe} if the recent, low values from \cite{Tang.Dauphas:2015,Tang.Dauphas.ea:2012} are assumed). This 11.8~\msun model is thus a plausible source for \iso{10}{Be} in the ESS.

A major uncertainty for the production of \iso{10}{Be} in CCSNe is the \bepa\ reaction rate.
The commonly used values for the rate from the JINA-REACLIB library \cite{Cyburt.Amthor.ea:2010}\footnote{\url{https://reaclib.jinaweb.org/index.php}} are estimates from \citet{Wagoner:1969}, based on very little experimental information.  Therefore, this rate is quite uncertain but it has been shown to have a significant impact on the \iso{10}{Be} yields \cite{Sieverding.Mueller.ea:2021}. 
Our study provides a thorough discussion of the sensitivity of \iso{10}{Be} production in CCSNe to the \bepa\ reaction rate, using a range of stellar progenitors and different models of the neutrino spectra. We further identify \bean\ as the next important reaction and discuss the role of the neutrino-induced reaction\benue.
Using the results of recent experiments that elucidate the lowest nuclear levels of the compound nucleus \iso{11}{B}, we derive an updated nuclear reaction rate for \bepa\ that is significantly higher than the rate commonly used in previous calculations. 
For the most plausible CCSN model, we show that the new rate reduces the \iso{10}{Be} yield significantly, making it insufficient to explain the \iso{10}{Be}/\iso{9}{Be} ratio in the ESS within the uncertainties of the reaction rate and CCSN neutrino emission.

This article is organized as follows: First, in Section \ref{sec:model} we briefly describe the 
supernova models and reaction network. In Section \ref{sec:yields} we first describe the general processes relevant for the nucleosynthesis of \iso{10}{Be} and discuss the sensitivity to the \bepa\ rate, as well as the role of the \bean\ and \benue\ reactions. 
In Section \ref{sec:data} we derive the new \bepa\ reaction rate and discuss the impact on the yields and implications for the ESS in Section \ref{sec:implications}. Finally we conclude in Section \ref{sec:conclusions}.

\section{\label{sec:model}Supernova model}

In the following we briefly describe the supernova models and nuclear reaction network used for this study.

To include the dependence on the stellar progenitor model, we use four CCSN and progenitor models that have been studied in Ref. \cite{Sieverding.Martinez.ea:2018}, covering stars with zero-age-main-sequence masses of 13~ \msun, 15~\msun, 20~\msun\ and 25~\msun\ and an initial composition of solar metallicity \cite{Lodders:2003}. Stellar evolution and the explosions have been calculated assuming spherical symmetry with the \textsc{Kepler} hydrodynamics and stellar evolution code \cite{Woosley.Heger.ea:2002,Weaver.Zimmerman.Woosley:1978}.
The explosions are driven by a parameterized piston tuned to yield explosion energies of $1.2\times 10^{51}\,\rm{erg}$. 

The models do not track the neutrino emission and we thus use the parameterization from Ref. \cite{Woosley.Hartmann.ea:1990} for the neutrino luminosity and assume the spectra to be constant. 
For the local neutrino flux at time $t$ and
radius $r(t)$ we assume 
\begin{equation}
    \Phi_\nu(t) =\frac{L_0}{4\pi\, r(t)^2}e^{-t/\tau_{\nu}}
\end{equation}
 with a timescale of $\tau_\nu=3\,\second$ and the luminosity $L_0$ adjusted to
 obtain a total energy of $3\times 10^{53}$ erg distributed equally among the six neutrino species.  
The spectra are assumed to be Fermi-Dirac distributions
with chemical potential $\mu_\nu=0$ and characterized with a constant spectral temperature $T_\nu$, related to the average neutrino energy as 
$\langle E_\nu \rangle \approx 3.15 \times T_\nu $. 
Since the neutrino spectra are a major uncertainty for the production of \iso{10}{Be} we explore two cases that generously cover the range of typical neutrino energies found in current simulations (e.g. \cite{Bollig.Yadav.ea:2020,Vartanyan.Burrows.ea:2019,Lentz.Bruenn.ea:2015}):
\begin{itemize}
    \item  High $E_\nu$, with $T_{\nu_e}=4\,\mev$, $T_{\bar{\nu}_e}=5\,\mev$, $T_{\nu_x}=T_{\bar{\nu}_x}=6\,\mev$
    \item  Low $E_\nu$, with $T_{\nu_e}=2.8\,\mev$, $T_{\bar{\nu}_e}=T_{\nu_x}=T_{\bar{\nu}_x}=4\,\mev$,
\end{itemize}
where $\nu_x$ ($\bar{\nu}_x$) represent $\mu$ and $\tau$ neutrinos (antineutrinos).

 In practice, the reaction network uses normalized, spectrum-averaged cross sections $\langle\sigma_\nu\rangle$ tabulated as a function of $T_\nu$ that are calculated from the energy-dependent cross section $\sigma_\nu(E_\nu)$ as:
  \begin{equation}
     \label{eq:spectrum_average}
     \langle \sigma_\nu \rangle (T_\nu)= \int\limits_0^{\infty} \sigma_\nu(E_\nu) f_\nu(E_\nu,T_\nu) dE_\nu,
 \end{equation}
 with the normalized distribution function 
 \begin{equation}
 f_\nu(E_\nu,T_\nu) = 
 \frac{N}{T_\nu^3}\frac{E_\nu^2}{1+\rm{exp}\left[E_\nu/T_\nu\right]},
 \end{equation}
  where $N=2/(3\zeta(3))\approx 0.55$.
  
In this study we also include the 11.8~\msun\ progenitor model from Ref. \cite{Banerjee.Qian.ea:2016} that has been used as the basis for 
a 3D CCSN simulation \cite{Mueller.Tauris.ea:2019}. 
For this progenitor we use a 1D explosion model from Ref. \cite{Sieverding.Mueller.ea:2021}, which has been tuned to give a much lower explosion energy of $0.2\times 10^{51}\,\rm{erg}$ that is consistent with the 3D simulation. We combine this model with the parameterized neutrino exposure described above and additionally include the neutrino emission from the simulation. 
 Note that we do not apply the corrections of Ref. \cite{Sieverding.Mueller.ea:2021}, which decrease $\langle E_{\bar{\nu}_e}\rangle$. The average neutrino energies predicted by the simulation are between the parameterized high $E_\nu$ and low $E_\nu$ cases and evolve with time. The simulation only covers the first 1.3~\second\ after core bounce and we extrapolate the neutrino luminosities as in \cite{Sieverding.Mueller.ea:2021} with an exponential decrease with $\tau_\nu=2\,\rm{s}$, and a linear decrease of the neutrino energies, reaching zero at 10~s after bounce. 

The explosion trajectories are post-processed with the open-source reaction network code \textsc{XNet}\footnote{\url{https://github.com/starkiller-astro/XNet}}, including about 2000 nuclear species, nuclear reactions from the JINA-REACLIB database \cite{Cyburt.Amthor.ea:2010} and most neutrino-nucleus reactions as in Ref.~\cite{Sieverding.Martinez.ea:2018}. The neutrino reactions on \iso{12}{C}, which are especially important for the production of \iso{10}{Be} are taken from shell-model calculations of Ref. \cite{Yoshida.Suzuki.ea:2008}. 

\iso{10}{Be} is produced by neutrino irradiation of \iso{12}{C}, mainly via the neutral current reaction \iso{12}{C}$(\nu_x,\nu_x' p p)^{10}$Be and the charged-current reaction \iso{12}{C}$(\bar{\nu}_e,e^+ n p)^{10}$Be. The charged current contribution makes the production sensitive to effects of neutrino flavor conversions \cite{Sieverding.Mueller.ea:2021,Yoshida.Suzuki.ea:2008}, which transform high energy $\bar{\nu}_x$ into $\bar{\nu}_e$, increasing the contribution of the charged-current channel. Similar effects have been discussed previously in the context of the CCSN $\nu$ process \cite{Kusakabe.Cheoun.ea:2019,Wu.Qian.ea:2015,Yoshida.Kajino.ea:2006}.
Therefore, in Section \ref{sec:implications} we also consider a complete swap of $\bar{\nu}_e \leftrightarrow \bar{\nu}_x$ to estimate the maximal effect of flavor transformations.  

The current version of the JINA-REACLIB library does not include a reaction rate for \iso{10}{Be}$(p,n)$\iso{10}{B} but
\citet{Kusakabe.Cheoun.ea:2019} have calculated the rate with the statistical model and they do not find a noticeable impact on the nucleosynthesis yields. Their values for the reaction rate are more than an order of magnitude lower than the \bepa\ reaction rate and we did not consider it in our calculations. 
\section{\label{sec:yields}Production of \texorpdfstring{\iso{10}{Be}}{10Be} and sensitivity study}
\subsection{\label{sec:overview}Overview}
 \begin{figure}
\includegraphics[width=\linewidth]{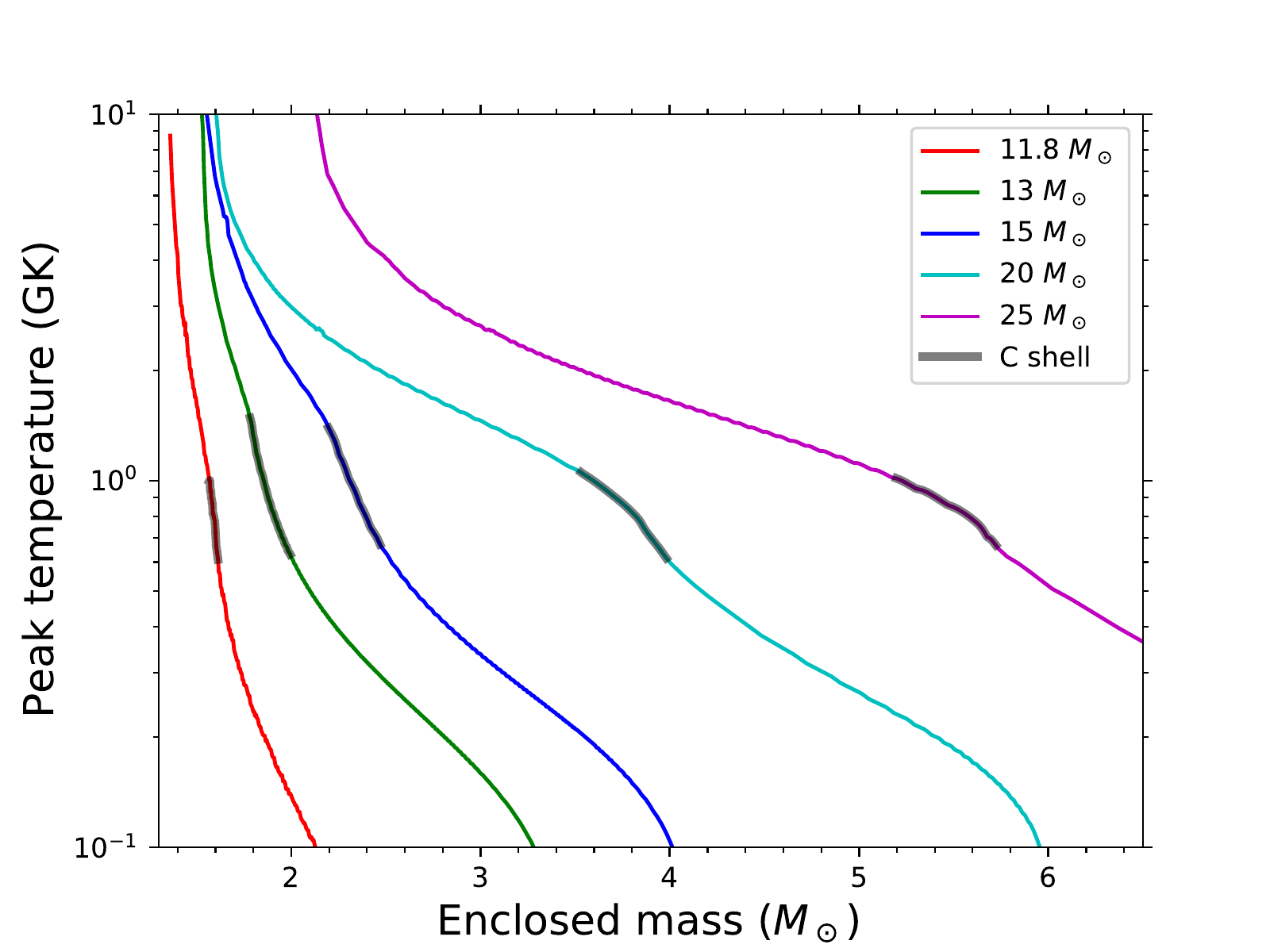}
\caption{\label{fig:tpeak} Peak temperature profiles for the supernova models used here. The C shell, where the production of \iso{10}{Be} occurs, is highlighted in gray. The peak temperature in the C shell ranges from 0.6 to 1.5~GK. }
\end{figure}
To understand the sensitivity of \iso{10}{Be} yields to the nuclear reaction rates, we illustrate in the following the processes that contribute to the production and destruction of \iso{10}{Be}.
As \iso{10}{Be} is produced from \iso{12}{C} via neutrino induced reactions, the initial mass
fraction ($X$) of \iso{12}{C} determines the parts of the CCSN ejecta where \iso{10}{Be} can be produced.
For the five CCSN models that we study here, the peak temperature in the C-shell ranges from 0.6~GK to 1.5~GK. This is illustrated in Fig. \ref{fig:tpeak} which shows the peak temperature for all the models with the C~shell highlighted as thick gray lines. 
In the following we describe the \iso{10}{Be} production for our calculations with the 15~\msun\ model in more detail as an example. 
The qualitative picture is similar for all the progenitor models.

\begin{figure}
\includegraphics[width=\linewidth]{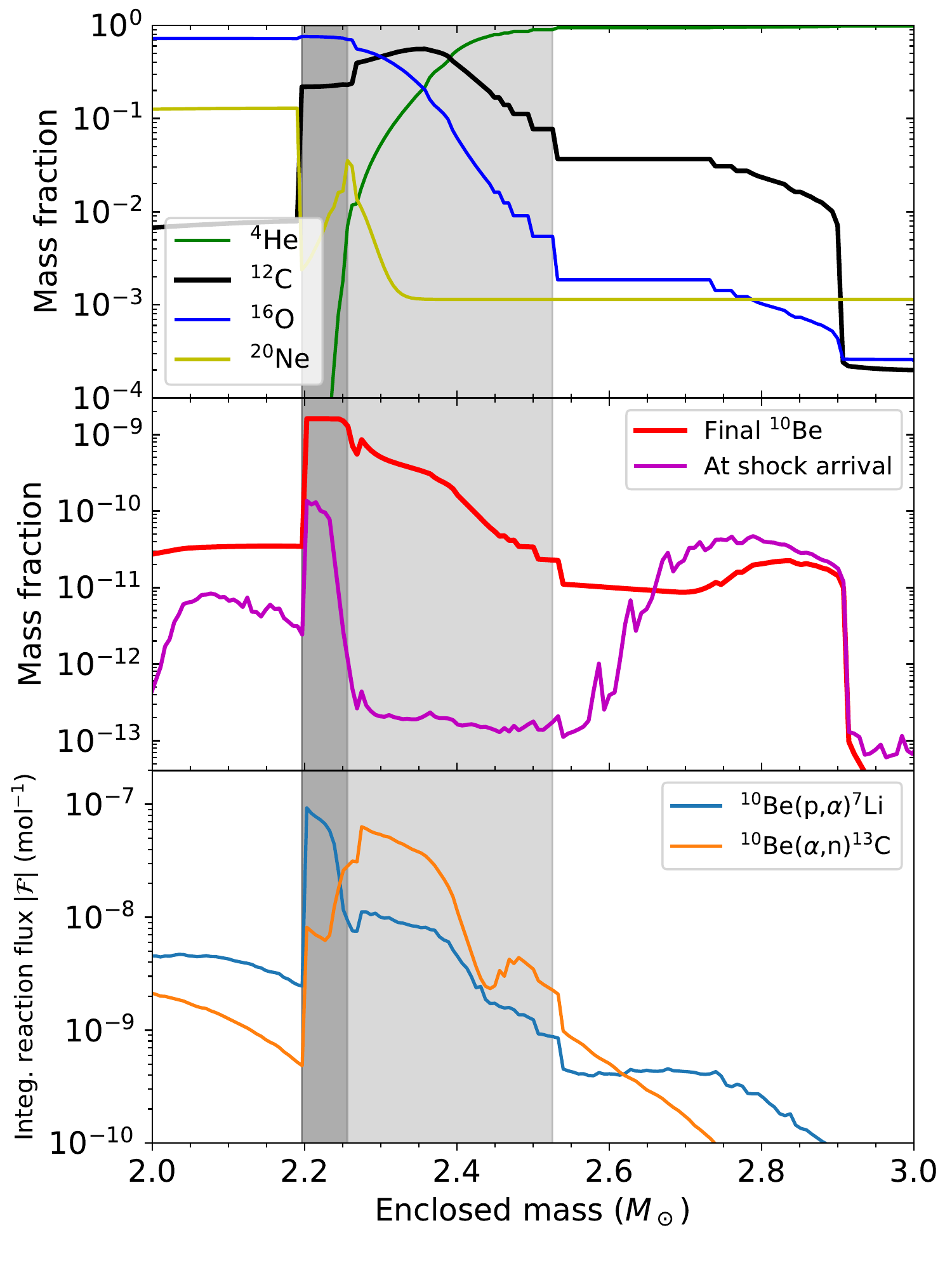}
\caption{\label{fig:fluxes} Mass fraction profile of the 15~\msun\ model. The top panel shows the pre-supernova mass fractions of several important isotopes to indicate the star's compositional layers. The middle panel shows the profiles of the \iso{10}{Be} mass fraction at 200~s (Final) and at the time of shock arrival, i.e., when the peak temperature is reached. Note that this time is different for each location. The bottom panel shows the absolute value of the integrated net reaction fluxes as defined in Eq. \eqref{eq:flux} due the two most important destructive reactions.}
\end{figure}

The top panel of Fig. \ref{fig:fluxes} shows the pre-supernova mass fraction profile of \iso{12}{C} as well as \iso{20}{Ne}, \iso{16}{O} and \iso{4}{He} for the 15~\msun\ progenitor model.
The O/Ne~shell below a mass coordinate of around $2.2\,\msun$ has been processed by C-shell burning, leading to a reduced mass fraction of \iso{12}{C} and an increase of \iso{20}{Ne}. 
The region with $X$(\iso{12}{C})$>0.05$, starting at a mass coordinate around $2.2\,\msun$\ (corresponding to a radius of $15,000\,\rm{km}$) is shown with a gray background and we will refer to this region as the C~shell in the following. The bottom of the C~shell is depleted in \iso{4}{He}, i.e., $X$(\iso{4}{He})$<0.01$, and this region is further shaded in dark gray.  
Above 2.25 \msun, $\alpha$ particles are leftover and in a narrow region $X$(\iso{12}{C})$>X$(\iso{16}{O}). The \iso{4}{He} mass fraction gradually increases toward the He~shell. The bottom of the He~shell, up to 2.9~\msun, also contains a low level of \iso{12}{C}.

The final mass fraction of \iso{10}{Be} at 200~s after the explosion is shown in the middle panel of Fig. \ref{fig:fluxes} and it is strongly peaked at the He-free bottom of the C shell. 
The bottom panel shows the time-integrated net reaction fluxes of the two dominant destruction reactions, 
\bepa\ and \bean. The integrated net flux of a reaction $A\rightarrow B$ is defined as 
\begin{equation}
\label{eq:flux}
    \mathcal{F}_{A\rightarrow B}=\int\limits_{t_0}^{t_{\rm{stop}}} \dot{y}_{A\rightarrow B}(t) + \dot{y}_{B\rightarrow A}(t) \;\rm{d}t,
\end{equation}
where the integral covers the whole duration of the calculation, i.e., 200~s, $\dot{y}_{A\rightarrow B}(t)$ and $\dot{y}_{B\rightarrow A}(t)$
are the change of the abundance of $A$ due to the forward and inverse reactions, respectively, at a given time.
The magnitude of the integrated flux of the destructive reactions is larger than the final abundances of \iso{10}{Be}, indicating that only a fraction of the produced abundance survives.
Fig. \ref{fig:fluxes} shows that we can distinguish between two regions that are indicated by the light and dark gray regions: 
\begin{enumerate}
    \item The bottom of the C shell, where \iso{4}{He} has been depleted almost completely,
   dominates the final yield of \iso{10}{Be} and \bepa\ is the most important destructive nuclear reaction.
    \item In the outer C~shell a significant mass fraction of \iso{4}{He} is still present and \bean\ is the dominant destructive reaction, limiting the contribution of this region to the production of \iso{10}{Be}.
\end{enumerate}

The \iso{10}{Be} yield is largest at the bottom of the C shell because at higher enclosed mass values, even though $X$(\iso{12}{C}) is higher, \iso{4}{He} becomes more abundant and therefore destroys \iso{10}{Be} via the \bean\ reaction. 
At the bottom of the C~shell, free protons need to be released first to inhibit the production of \iso{10}{Be}.
In addition to nuclear reactions and 
photo-disintegration activated by the shock, neutrino-induced reactions, such as the neutral current reaction $^{16}$O$(\nu_x,\nu_x'p)^{15}$N, play an important role as a proton source. In particular at later times, when the temperature is too low for photo-disintegration, neutrino reactions continue to provide protons. The neutrino reactions are also important proton sources in the outer C~shell and the bottom of the He~shell, where the peak temperatures are low.
Before the shock hits, the temperature is relatively low, allowing the neutrino interactions to build up \iso{10}{Be}.
When the shock hits,
the destruction reactions are most effective while the temperature remains high. 
Free protons are quickly consumed, suppressing the \bepa\ reaction and allowing the \iso{10}{Be} abundance to recover. Due to neutrino-interactions, the proton abundance also increases again. In cases where the temperature remains high for long enough or when the reaction rate is enhanced, the destructive reaction becomes active again, leading to a decrease of the \iso{10}{Be} mass fraction at late times. This possibility of late-time destruction is important for the impact of an enhanced reaction rate as illustrated in Fig. \ref{fig:evol_example} and is discussed in more detail below.
\begin{figure}
\includegraphics[width=\linewidth]{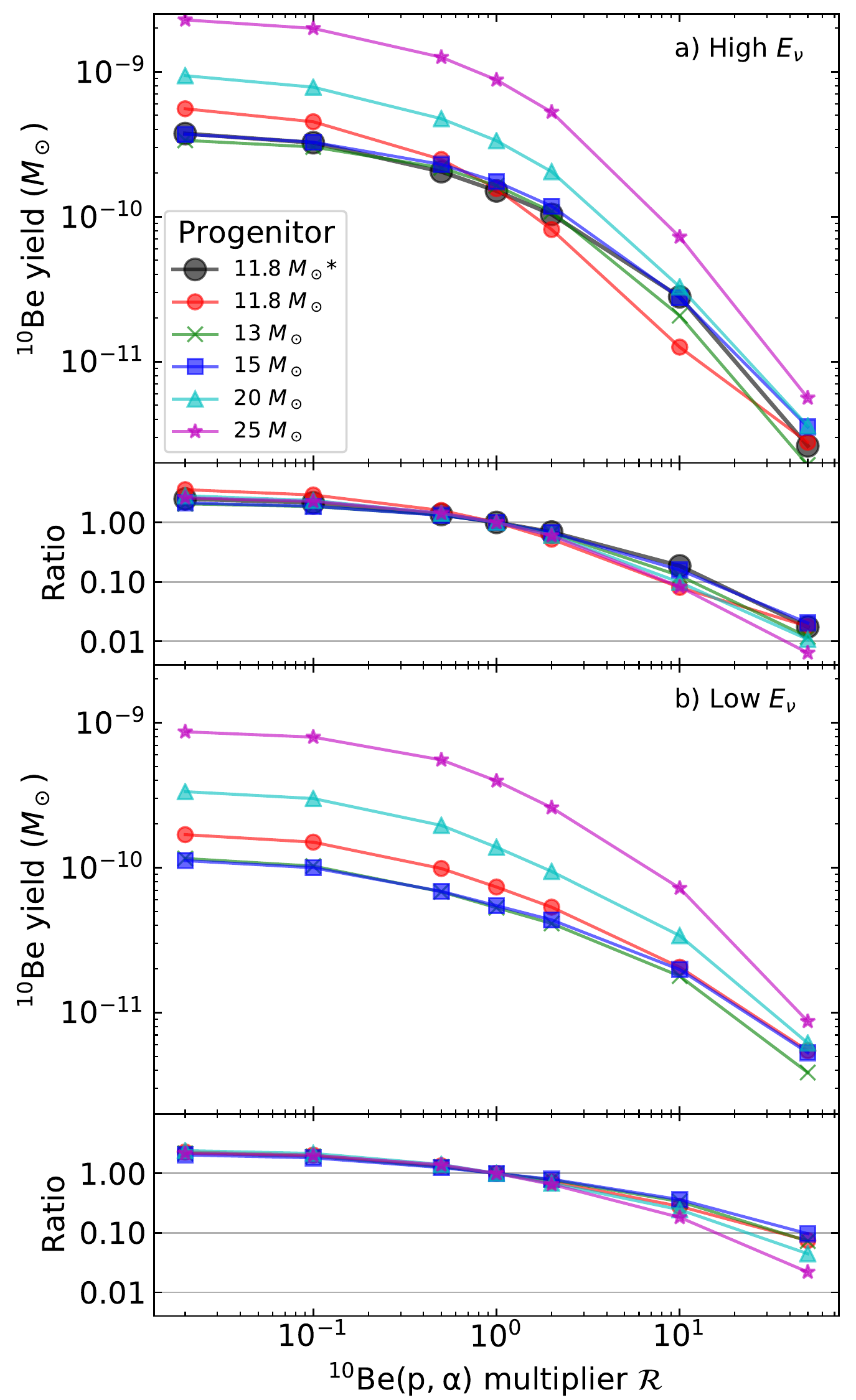}
\caption{\label{fig:sensitivity} Impact of the variation of the \iso{10}{Be}$(p,\alpha)$ reaction rate on the \iso{10}{Be} yields and ratio relative to the result with the JINA-REACLIB reaction rate. The upper two panels are for high $E_\nu$ and the lower two panels for low $E_\nu$. The results with the neutrino spectra from the 3D simulation of the $11.8$ \msun\ model is also included in the upper panel ($11.8$ \msun*).}
\end{figure}
Most of the 
\iso{10}{Be} yield is only produced after the shock has passed. This is illustrated by the middle panel of Fig. \ref{fig:fluxes}, which shows that the \iso{10}{Be} mass fraction at the time when the peak temperature is reached is much smaller than the final mass fraction everywhere in the C~shell.
At the bottom of the He~shell, however, where \iso{12}{C} is also present, the peak temperature is too low for the efficient destruction of the SLR by the SN shock. Here, the final \iso{10}{Be} mass fraction is lower than the mass fraction at the time of shock arrival, illustrating that the $(p,\alpha)$ reaction continues to be active after the shock has passed.

The relative contribution of the bottom of the C~shell varies with progenitor mass due to different sizes of the shells, different temperatures and distance from the proto neutron star as well as differences in the initial composition. For the 11.8~\msun\ model about $50\,\%$ of the total \iso{10}{Be} yield is provided by the He-free bottom of the C~shell. For the 15~\msun\ model the contribution of the bottom of the C~shell is almost $70\,\%$.

In the 25~\msun\ model, the O/Ne shell, which is much larger in terms of mass than the C~shell, has a \iso{12}{C} mass fraction that is higher by almost an order of magnitude compared to the 15~\msun\ model. Therefore, for the 25~\msun\ model, the O/Ne shell contributes $80\%$ of the \iso{10}{Be} yield.

\subsection{Sensitivity to \texorpdfstring{\bepa}{bepa}}
\label{sec:bepa}

To explore the sensitivity of the \iso{10}{Be} yield to the \bepa\ rate, we perform post-processing reaction network calculations including a global, i.e., temperature-independent, multiplier $\mathcal{R}$ for this rate.
The existence of resonances can have a large impact, as we will show in Section \ref{sec:implications}. Therefore, we explore a large range, increasing and decreasing the rate by factors 2, 10 and 50 relative to the JINA-REACLIB value. 
The inverse reaction, \iso{7}{Li}$(\alpha,p)$, is strongly suppressed due to the negative Q-Value and Coulomb barrier. Furthermore, the mass fraction of \iso{7}{Li} does not exceed $10^{-7}$, making it a relatively rare target. Thus, for the conditions relevant here, we assume the reverse reaction to be negligible and keep its rate unchanged for the sensitivity analysis. The effect of the inverse reaction is included in Section \ref{sec:implications}.

Fig. \ref{fig:sensitivity} shows the \iso{10}{Be} yield as a function of the reaction rate multiplier for the five stellar models considered here. The top panel shows the results for high $E_\nu$ and the bottom panel the results for low $E_\nu$.
The neutrino spectra from the 11.8~\msun simulation are time-dependent, but tend to be in between the high and low energy cases and the results are included in the top panel.

For a suppression of the \bepa\ reaction, i.e., for $\mathcal{R}<1$, the yields moderately increase.
The increase is limited by the \bean\ reaction, which eventually becomes dominant. The bottom panel of Fig. \ref{fig:fluxes} shows that the reaction flux through \bepa\ is at most one order of magnitude higher than the flux through \bean. The latter reaction can thus be expected to become dominant if $\mathcal{R}\lessapprox 0.1$. Accordingly, in Fig. \ref{fig:sensitivity} we only see a small increase of the yield from $\mathcal{R}=0.1$ to $\mathcal{R}=0.02$.

For the 11.8~\msun\ model with high $E_\nu$,
the increase relative to the results with the default reaction rate is largest, reaching a factor of 3.5 for $\mathcal{R}=0.02$.
For all other models, the ratio of the yield relative to the yield with the default reaction rate ranges from a factor of $2.0$ to $2.7$.
Ref. \cite{Sieverding.Mueller.ea:2021} found that turning off the \bepa\ reaction completely increases the yield by a factor 3, which is consistent with the range of values we find here.

\begin{figure}
    \centering
    \includegraphics[width=\linewidth]{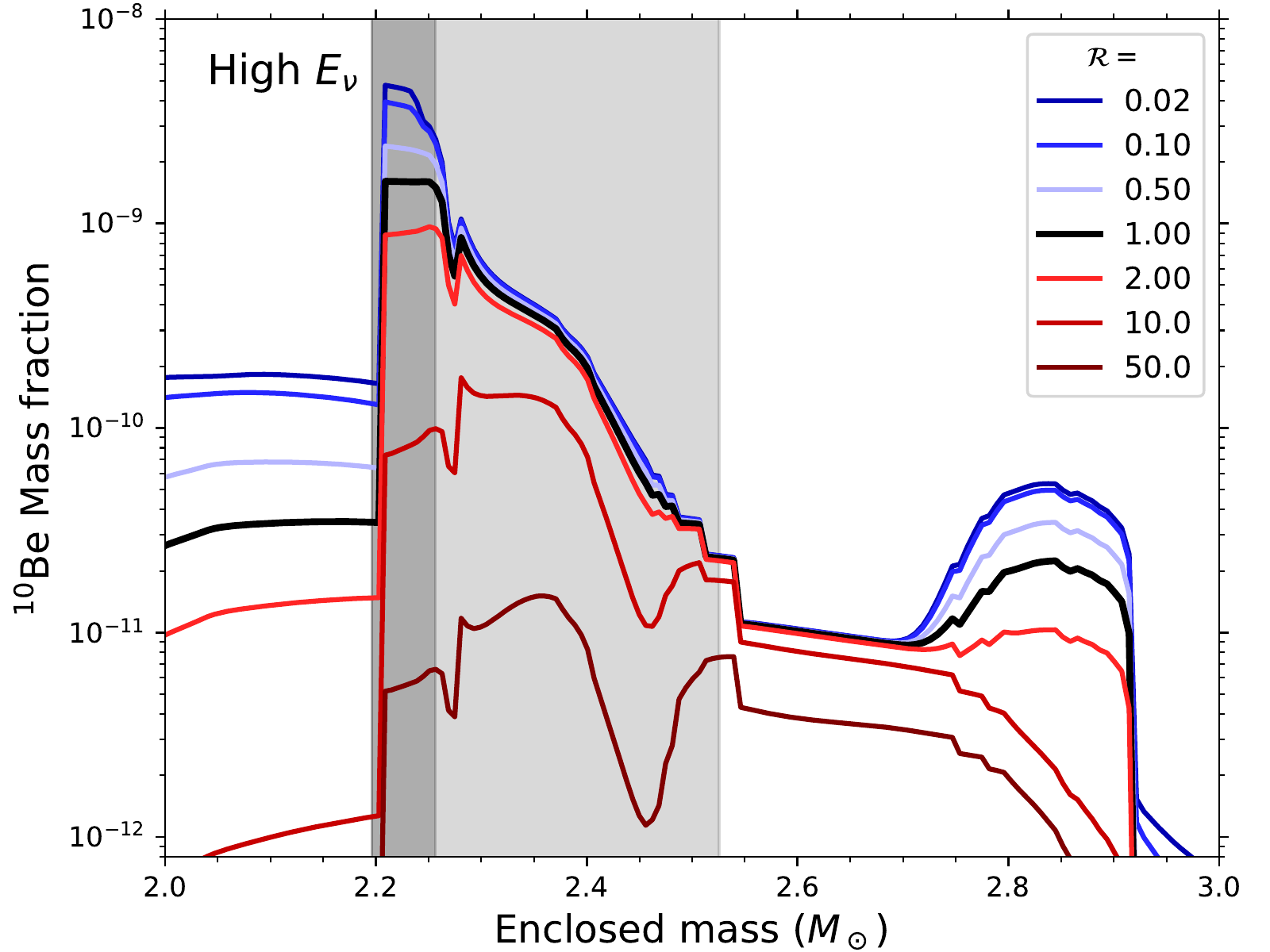}
    \caption{Mass fraction profile of \iso{10}{Be} for the 15~\msun model with high $E_\nu$ for the range of values of the \bepa reaction rate multiplier $\mathcal{R}$. The contribution of the bottom of the  C~shell decreases with increasing $\mathcal{R}$.}
    \label{fig:profile_sensitivity}
\end{figure}

An increase of the \bepa\ rate, i.e., $\mathcal{R}>1$ in Fig. \ref{fig:sensitivity}, efficiently reduces the yield. As explained above, the \bepa\ reaction is still active after the shock has passed and reduces the peak mass fraction of \iso{10}{Be}. Given a sufficiently high temperature and high rate, the reaction will tend to destroy practically all of the \iso{10}{Be}.
For an individual trajectory, an increase of $\mathcal{R}$ reduces the peak \iso{10}{Be} mass fraction that is accumulated after the shock and it also extends the time during which the destruction operates after the production has ceased. 
This makes the final yield very sensitive to $\mathcal{R}$. Figure \ref{fig:profile_sensitivity} illustrates the strong impact of the enhancement of  the rate and it also shows that the effect is most pronounced at the bottom of the C~shell, where the temperature is highest. For $\mathcal{R}=10$, only $8-15\,\%$  ($18-35\,\%$) of the yield with the default rate are left, with high (low) $E_\nu$. In relative terms, the reduction is largest for the 25~\msun\ model because of the large contriution from the O/Ne shell and the higher peak temperature which enhances the reaction.

As $\mathcal{R}$ increases, \iso{10}{Be} is more effectively destroyed at high temperatures, shifting the main production to lower temperatures. Furthermore, the \bepa\ reaction also remains effective at a lower temperature and thus at later times, when protons from thermonuclear reactions are increasingly rare. Both of these effects make the role of neutrino-induced reactions as proton sources more important for larger $\mathcal{R}$. As a consequence, the yields decrease faster with $\mathcal{R}$ for high $E_\nu$ than for low $E_\nu$, as shown by the ratios in Figure \ref{fig:sensitivity}. 
 \begin{figure}
\includegraphics[width=\linewidth]{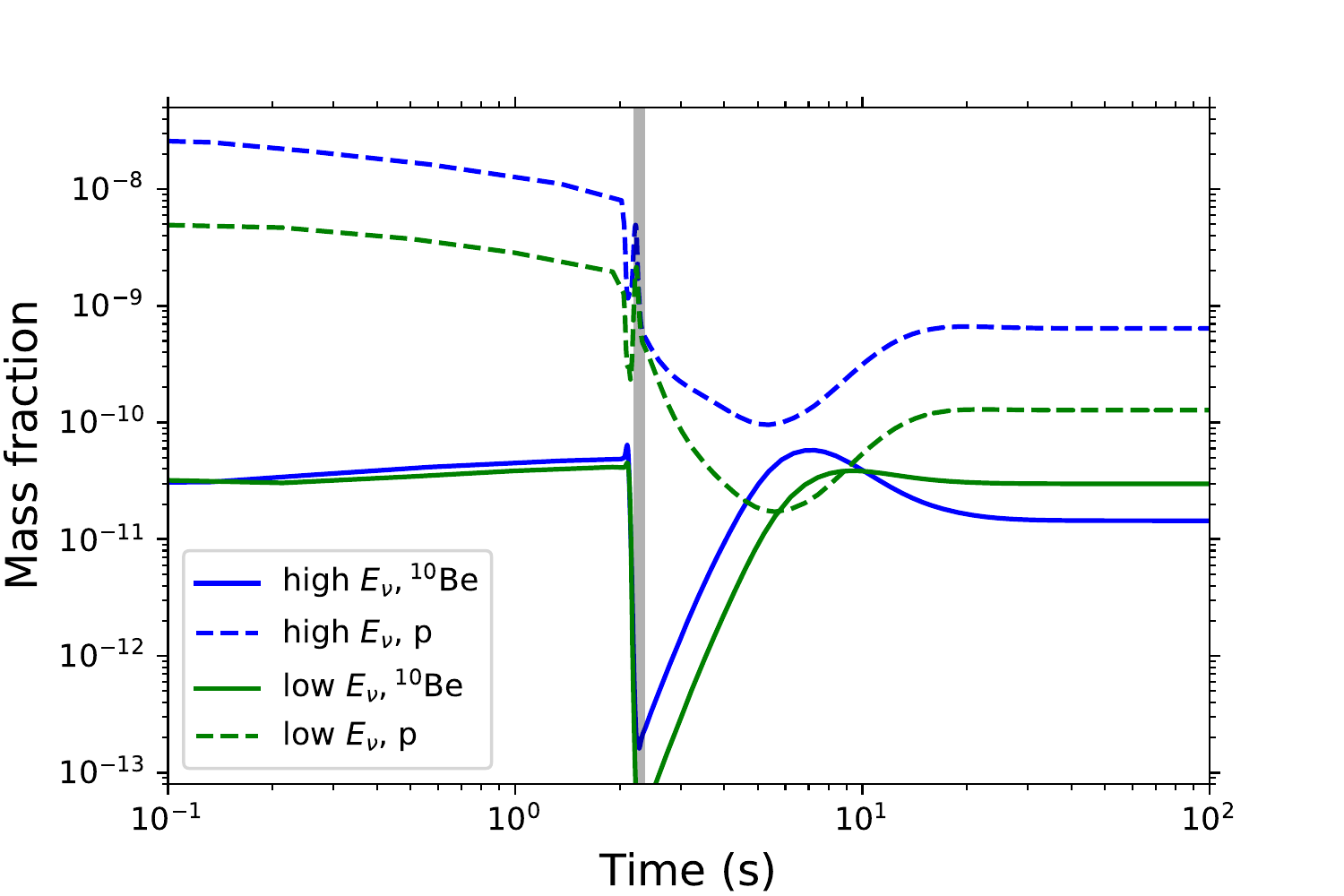}
\caption{\label{fig:evol_example} Time evolution of the mass fractions of free protons and \iso{10}{Be} for the 15~\msun\ model in the outer C shell at mass coordinate of 2.32~\msun\ and with the \bepa\ rate enhanced by a factor 50 for low and high $E_\nu$. With high $E_\nu$ the proton mass fraction is enhanced by almost factor of 10, leading to a lower final \iso{10}{Be} mass fraction than in the low $E_\nu$ case.}
\end{figure}

The role of neutrinos as a proton source also leads to the effect that
the \iso{10}{Be} yields with high $E_\nu$ become lower than the
results with low $E_\nu$ for higher values of $\mathcal{R}$, which is
visible when comparing the upper and lower panel of
Fig. \ref{fig:sensitivity}.  This effect is more clearly illustrated in
Fig. \ref{fig:evol_example} which shows the time evolution of the
\iso{10}{Be} and proton mass fractions for a zone at 2.32~\msun~ in
the 15~\msun\ model for $\mathcal{R}=50$ comparing the high $E_\nu$ and low
$E_\nu$ cases.  The arrival of the SN shock is marked as a gray vertical line
when most of the initially produced \iso{10}{Be} is destroyed. The
final mass fraction is determined by the production after the shock has
passed. With high $E_\nu$ the proton mass fraction is significantly
higher, allowing the enhanced \bepa\ reaction to reduce the
\iso{10}{Be} mass fraction even at 10~s, when the temperature has
dropped to 0.4~GK. As a result, even though the peak mass fraction of
\iso{10}{Be} reached at 6~s is higher for high $E_\nu$, the final mass
fraction is higher for low $E_\nu$ because of the lower proton
abundance.  Note that this inversion does not occur with the default
\bepa\ reaction rate, because the reaction is too slow at low
temperatures.  For the 15~\msun\ model, this effect becomes only
noticeable for $\mathcal{R}=50$, when the yield is
$3.5\times 10^{-12}\,\msun$ for high $E_\nu$ and
$5.3\times 10^{-12}\,\msun$ with low $E_\nu$.

The effect is more pronounced for the 11.8~\msun\
model because it has a much lower explosion energy and peak
temperatures are generally lower, increasing the importance of neutrinos to provide protons.  
For example, for the 11.8~\msun\ model with $\mathcal{R}=10$, the high
$E_\nu$ case actually results in the lowest yield among all models for this
value of $\mathcal{R}$, which is $1.3\times 10^{-11}\,\msun$. For
comparison, the yield is $2.8\times 10^{-11}\,\msun$ with the neutrino
spectra from the simulation and $2.1\times 10^{-11}\,\msun$ with low
$E_\nu$.  The yield with the neutrino spectra from the simulation is
highest, because the neutrino temperatures decrease with time and the
luminosities decrease faster, reducing the proton mass fraction at
late times.   This illustrates the complex interplay between the
sensitivity to the reaction rate and the sensitivity to the neutrino
spectra in the $\nu$ process.

In Section \ref{sec:overview} we have shown that the He-free bottom of the C~shell dominates the yields for the JINA-REACLIB reaction rate. 
Since the \bepa\ reaction is the main destructive channel in that region, its relative contribution to the total yield decreases first as $\mathcal{R}$ increases. This is illustrated in Fig. \ref{fig:profile_sensitivity} for the 15~\msun model and shows that the He-rich, outer C~shell already becomes the dominant production region for $\mathcal{R}=10$.
For different progenitors the differences in the relative contributions of the different shells and the different temperatures in the respective regions determine the details of the behaviour of the yield as a function of $\mathcal{R}$. The overall trend, however, is very similar for all the models, with a slight increase of the yield for $\mathcal{R}<1$ and a much steeper decrease for $\mathcal{R}>1$.

\subsection{Sensitivity to \texorpdfstring{\bean}{10Be(a,n)}}
The reaction \bean\ has already been mentioned in Ref. \cite{Yoshida.Suzuki.ea:2008} as most important destructive reaction using, however, an older and much higher value of the rate. With the currently recommended rate in the JINA-REACLIB library we find it to be the second most important destruction channel for \iso{10}{Be} after \bepa.
Fig. \ref{fig:fluxes} shows that \bean\ is generally dominant in the outer regions, where $\alpha$ particles are leftover from incomplete He-burning. The reaction limits the contribution from this region to the total \iso{10}{Be} yield. The default reaction rate in the JINA-REACLIB database is up to a factor 100 lower than the values in STARLIB \cite{Sallaska.Iliadis.ea:2013}.
Due to this large difference, we explore a global variation of the reaction rate from the JINA-REACLIB database by factors 10 and 50 up and down, while keeping the \bepa\ reaction rate from the JINA-REACLIB library.
Increasing the \bean\ reaction rate by a factor 50 (10) for the 15 \msun\ model with high $E_\nu$ reduces the yield from $1.7\times 10^{-10}~\msun$ by a factor 1.9 (1.4) to $0.9\times 10^{-10}~\msun$ ($1.2\times 10^{-10}~\msun$). Suppressing the reaction rate by a factor 50 (10), on the other hand yields an increase by a factor 1.6 (1.4), to $2.7\times 10^{-10}~\msun$  ($2.3\times 10^{-10}~\msun$) of \iso{10}{Be}. Based on this, we estimate the overall uncertainty due of the \iso{10}{Be} yields due to this reaction rate to be up to a factor of~2. 
Thus, for a full understanding of the \iso{10}{Be} production in CCSNe, a re-evaluation of experimental constraints on the cross section for \bean\  or a direct measurement is highly desirable.

\subsection{Sensitivity to \label{sec:benue}\texorpdfstring{\benue}{10Be(nue,e-)10B}}

The reaction \benue\ may contribute to the destruction of \iso{10}{Be}
\cite{Sieverding.Mueller.ea:2021}. An estimate of the cross section
was previously provided in the global compilation of neutrino cross
sections of Ref.~\cite{Sieverding.Martinez.ea:2018}. Here, we improve
it by combining shell-model calculations and experimental data. From
the beta-decay data of the mirror nucleus of \iso{10}{Be},
\iso{10}{C}, we expect that $\nu_e$ absorption on \iso{10}{Be} will
have a large Gamow-Teller (GT) contribution to the $1^+$ state at
718~keV in \iso{10}{B}, $\log ft=3.0426(7)$, $B(GT)=5.573(9)$
\cite{Tilley04}. In addition, we expect a Fermi contribution to the
$0^+$ state at 1.740~keV. To determine whether other states may
further enhance the cross section, we have performed a shell model
calculation with the code NATHAN
\cite{Caurier.Martinez-Pinedo.ea:2005} employing the Cohen-Kurath
interaction \cite{Cohen65}. The theoretical GT matrix elements have
been reduced by a quenching factor
$q=0.82$~\cite{Chou.Warburton.Brown:1993}.  We find that the GT
transition at 718~\kev dominates the cross-section at the
relevant neutrino energies. No noticeable reduction of the
\iso{10}{Be} yield is found, even assuming $T_{\nu_e}=\,6\,\mev$,
because the abundance of \iso{10}{Be} is never large enough to allow
for a noticeable number of interactions. For completeness, we provide
the values of the spectrum averaged cross-sections in
Table~\ref{tab:be10nuabs}.

\begin{table}[hbt]
  \caption{\label{tab:be10nuabs}
Spectrum averaged cross-section for $\nu_e$ absorption on
  \iso{10}{Be} as defined in Eq. \eqref{eq:spectrum_average}.} 
\begin{ruledtabular}
    \begin{tabular}{lrrrrrrr}
    $T_\nu$ (MeV) & 2.8 & 3.5 & 4.0 &5.0 & 6.4 & 8.0 & 10.0 \\
    $\langle\sigma_{\nu_e}\rangle (10^{-42} \rm{cm}^2)$  &  14.6  & 22.8 &  29.7 & 46.2 &73.2 & 105.3 & 141.3 \\
    \end{tabular}
\end{ruledtabular} 
\end{table}

\section{\label{sec:data}New \texorpdfstring{\bepa}{10Be(p,a)} reaction rate}

\begin{figure}[b]
\includegraphics[width=\linewidth]{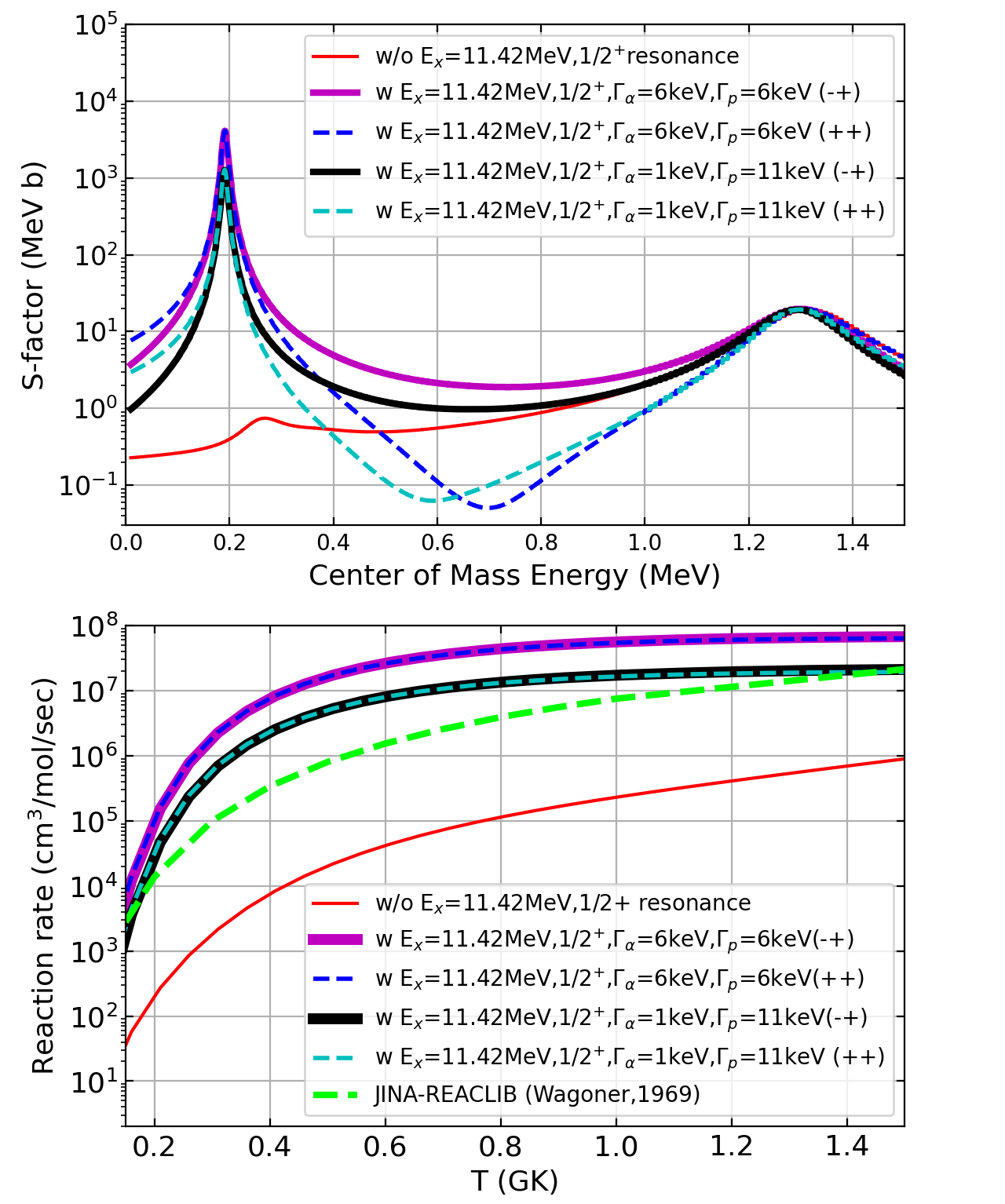}
\caption{\label{fig:rate} $R$-matrix calculations to assess the impact of the 193 keV 1/2$^{+}$ resonance on the \iso{10}{Be}$(p,\alpha)$ reaction cross section. The upper panel shows the $S$-factor and the lower panel the reaction rate as a function of temperature. Results without the 193~keV 1/2$^{+}$ resonance are shown (red line) as well several results assuming different resonance widths and assumptions about the interference with the  3/2$^{+}$ resonance (see text). For comparison, the rate from \cite{Wagoner:1969} is also shown (green dashed line). The result with $\Gamma_{p}$~=~6~keV, $\Gamma_{\alpha}$~=~6~keV, (+-) is the new recommended rate.} 
\end{figure}

The \bepa\ reaction proceeds through the \iso{11}{B} compound nucleus where states above the proton emission threshold play an important role. Current estimates of the \bepa\ reaction rate are based on estimates by \citet{Wagoner:1969} and it is the default reaction rate in JINA-REACLIB. We refer to this rate as the REACLIB rate hereafter. While the REACLIB rate was based on a constant S-factor approximation, this reaction is expected to be dominated by isolated resonances in the Gamow window. In recent years, states near the proton threshold in \iso{11}{B} have gained significant attention as they can explain the puzzling $\beta^{-}$p$^{+}$ decay in \iso{11}{Be}. A recent experiment, which directly measured the protons and their energy distribution, shows that the decay proceeds sequentially though a narrow resonance [$E$~=~11425(20)~keV, $\Gamma$~=~12(5)~keV, J$^{\pi}$~=~(1/2$^{+}$, 3/2$^{+}$)] in \iso{11}{B}~\cite{Ayyad19}. Preceeding this experiment, it was shown that Shell Model Embedded in the Continuum (SMEC) calculations strongly favor the J$^{\pi}$=1/2$^{+}$ assignment over 3/2$^{+}$~\cite{Okol20}. This 1/2$^{+}$ resonance at $\approx$193~keV (\iso{11}{B} $S(p)$~=~11228~keV), proves to be most crucial for the \bepa\ reaction rate as it is well within the Gamow window and provides the dominant contribution to \bepa\ reaction rate throughout, as discussed in detail below.

In this work, we have re-evaluated the \bepa\ reaction rate  using the $R$-matrix approach using the code \texttt{AZURE2}~\cite{azure,azure2}. For the $R$-matrix calculations, other than the known energy levels in \iso{11}{B} above the proton emission threshold, we have included the recently observed 11.425 MeV state (193~keV 1/2$^{+}$ resonance, discussed above), which is one of the main highlights of this work. In addition, we have also included a 3/2$^{+}$ state at 11.490 MeV, predicted by \citet{Jonas19}  to explain the $\beta$-delayed $\alpha$-spectrum from \iso{11}{Be} and later discussed by \citet{Okol20}.\\ All energy levels considered for the current $R$-matrix calculations and corresponding parameters are shown in Table~\ref{tab:level_params}. Energies of the states and partial widths are adopted from the NNDC database wherever possible \cite{NNDC}. In the absence of data, reduced widths are adopted as $\approx$0.01$\gamma^{2}_{W}$, where $\gamma^{2}_{W}$ is the Wigner limit \cite{ANGULO19993}.  The results of the $R$-matrix calculations are shown in Fig.~\ref{fig:rate} where the upper panel shows the astrophysical $S$-factor as a function of center-of-mass energy and the lower panel shows the reaction rate versus temperature. 
Fig.~\ref{fig:rate} (upper panel) shows the $S$-factor with and without the 193 keV resonance contribution. The cross section without the 193 keV resonance (red line) is dominated  in the Gamow window (90 keV-410 keV) by the low energy tail of the broader, higher lying 1/2$^{+}$ resonance ($E_x$~=~12.55~MeV). When the 193~keV 1/2$^{+}$ resonance is included (magenta and blue lines), the cross section in the Gamow window is completely dominated by this resonance and is several orders of magnitude higher. Moreover, the effect of a 3/2$^{+}$ state at 11.490~MeV, which is included in the calculations, is negligible for the overall cross-section. We also considered the interference between the 1/2$^{+}$ states where (-+) and (++) represents the different relative signs of channels and found that this did not have a significant effect on the reaction rate.  

Fig.~\ref{fig:rate} (lower panel) shows the comparison of new reaction
rates, with and without the 193 keV 1/2$^{+}$ state, as well as  to th REACLIB rate \cite{Cyburt.Amthor.ea:2010} (green dotted line) in the temperature range relevant for  the production of \iso{10}{Be} as indicated in Fig. \ref{fig:tpeak}.
A maximum predicted rate (Magenta line) is obtained when the 193 keV 1/2$^{+}$ resonance is included with its maximum strength ($\Gamma_{p}$~=~6~keV, $\Gamma_{\alpha}$~=~6~keV). Here, it is worth mentioning that from the recent resonant elastic scattering experiment \iso{10}{Be}(p,p),
which confirms the presence of the 193 keV 1/2$^{+}$ resonance,
R-matrix calculations reproduce the data well only if
$\Gamma_{p}$~=~6~keV, $\Gamma_{\alpha}$~=~6~keV
\cite{Mittig_Yassid}. Therefore, these widths are based on recent
experimental results and we refer to this rate as the new recommended rate in the following. To
assess the sensitivity of the reaction rate to the partial widths of the resonance at 193 keV, we also considered another extreme. The lower predicted rate
(black line), which includes the 1/2$^{+}$ resonance but with a
smaller strength resulting from $\Gamma_{p}$~=~11~keV and
$\Gamma_{\alpha}$~=~1~keV. This rate hereafter is referred as minimum rate. The difference between our recommended reaction rate and the minimum rate is what we refer to as the current uncertainty in the \bepa\ reaction rate. The minimum rate and the recommended rate are factors of $\sim$200 to $\sim$1000 higher compared to the rate derived without the 193 keV 1/2$^{+}$ resonance, respectively.  This shows that the 193 keV resonance has a large impact on the \bepa\ reaction rate in the Gamow window relevant for CCSNe. Compared to the REACLIB rate, our minimum and recommended rate are higher by a factor of $\approx$6 and $\approx$20 in the Gamow window.

\begin{table}
\centering
\caption{Levels included in the $R$-matrix calculations.}
\label{tab:level_params}
\begin{ruledtabular}
\begin{tabular}{ ccl }
Energy (MeV)  &   J$^{\pi}$   & partial widths (keV) \\ \hline
11.272  &  9/2$^{+}$  &  $\Gamma_{p}$=10$^{-15}$, $\Gamma_{\alpha}=110$\\ 
11.425   & 1/2$^{+}$  &  $\Gamma_{p}$=6,11, $\Gamma_{\alpha}$=6,1 \\ 
 11.490    &  3/2$^{+}$  &  $\Gamma_{p}$=10$^{-4}$ $\Gamma_{\alpha}$=93 \\ 
 11.600         &  5/2$^{+}$ &   $\Gamma_{p}$=10$^{-5}$, $\Gamma_{\alpha}$=90, $\Gamma_{n}$=90                     \\ 
 11.893        &   5/2$^{-}$         &  $\Gamma_{p}$=10$^{-4}$,  $\Gamma_{\alpha}$=100, $\Gamma_{n}$=94                 \\ 
 12.040        &   7/2$^{+}$  &  $\Gamma_{p}$=10$^{-3}$, $\Gamma_{\alpha}$=500, $\Gamma_{n}$=500                    \\ 
 12.550        &  1/2$^{+}$    &  $\Gamma_{p}$=100, $\Gamma_{\alpha}$=105                  \\ 
\end{tabular}
\end{ruledtabular}
\end{table}

For the nucleosynthesis calculations in Section \ref{sec:implications}, we consider both, the recommended and minimum rates to estimate the impact of the remaining uncertainty in \bepa\ reaction rate. 

\section{\label{sec:implications}
Impact of the new \texorpdfstring{\bepa}{10Be(p,a)} reaction rate}
\subsection{Impact on Supernova Yields }
The new \bepa\ reaction rate is  based on the updated nuclear data as described in Section \ref{sec:data}. 
Our recommended rate is factor of $\sim$20 higher compared to the REACLIB rate. The reaction network calculations use the parameterization given in Appendix \ref{sec:reaclib_fit} and include the rate for the inverse reaction,\liap, using detailed balance. The inverse reaction is only noticeable with high $E_\nu$ and results in an increase in the yield by less than 20\%.

\begin{figure}
    \centering
    \includegraphics[width=\linewidth]{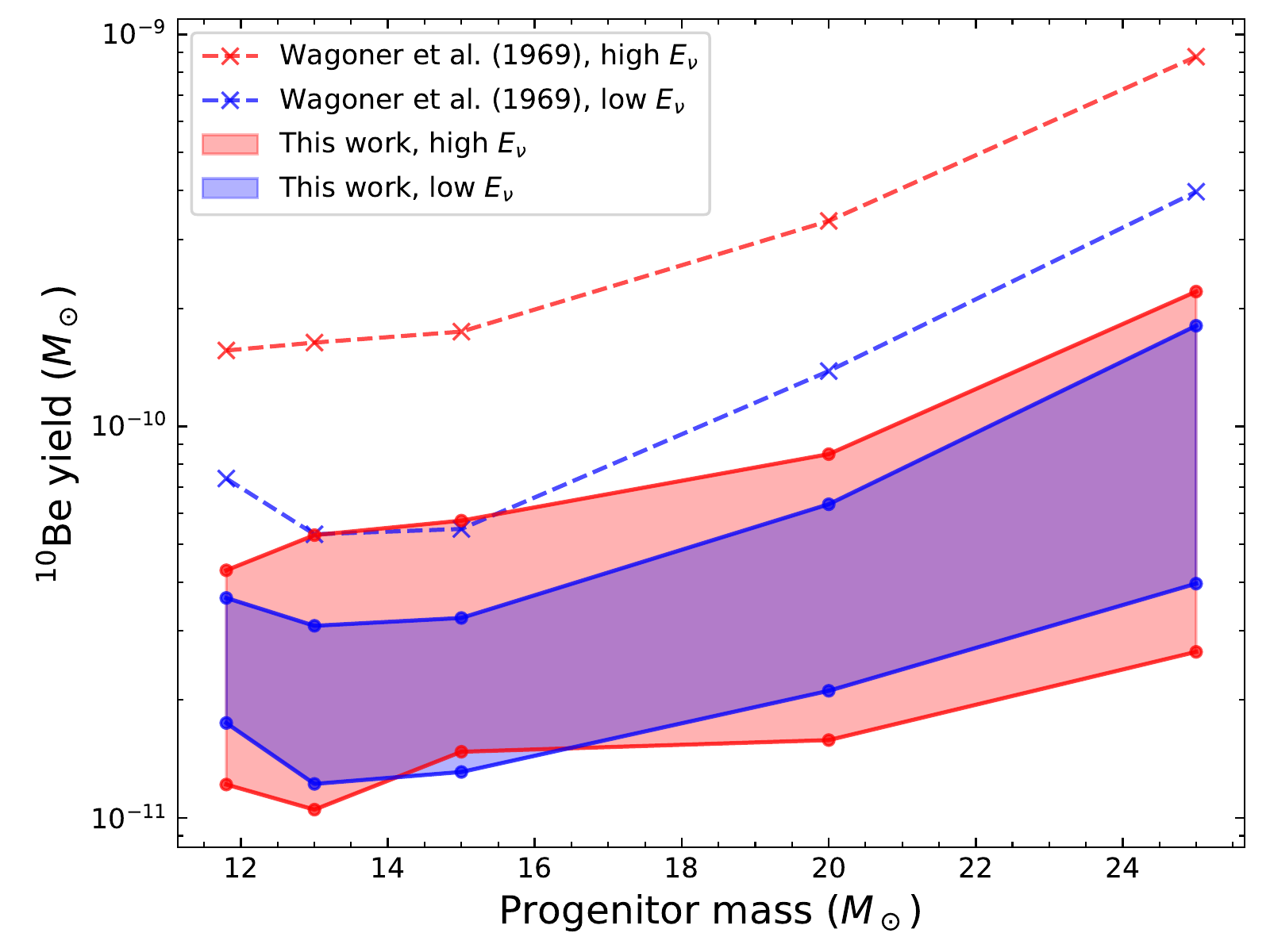}
    \caption{\iso{10}{Be} yields for the models studied with high and low $E_\nu$. The red and blue shaded bands indicate the range of results between the minimum and recommended rate from Section \ref{sec:data}. For comparison, the dashed lines indicate the results with the 
    JINA-REACLIB rate from \cite{Wagoner:1969} from models with the corresponding neutrino spectra. Note that we assumed and explosion energy of $0.2\times 10^{51}$~erg for the 11.8~\msun and $1.2\times 10^{51}$ erg for all other models.}
    \label{fig:new_yields}
\end{figure}

Figure \ref{fig:new_yields} summarizes the range of \iso{10}{Be} yields obtained with the new reaction rates for the five progenitor models and compares the high and low $E_\nu$ cases. The results with the REACLIB rate are shown for comparison. 
From Figure \ref{fig:new_yields}, the \iso{10}{Be} yield trend shows that the yield increases with increasing progenitor mass from 15~\msun to 25~\msun because the amount of material in the C shell increases.
As noted by \cite{Banerjee.Qian.ea:2016}, however, the yields for the progenitors from 11.8~\msun to 15~\msun are relatively similar. For high $E_\nu$ the yield even decreases slightly from 11.8~\msun to 13~\msun. This is because the 11.8~\msun model assumes a lower explosion energy, leading to lower peak temperatures that favor the survival of \iso{10}{Be}.
The spread in the \iso{10}{Be} yields  with the minimum and recommended rate for a given $E_\nu$ model is indicated by the colored bands and represents the impact of current uncertainty in the \bepa\ reaction rate. This band is slightly narrower with low $E_\nu$ compared to high $E_\nu$ due to the role of neutrino-induced reactions as proton sources.
Compared to the results with the  REACLIB rate, using the new recommended reaction rate and high $E_\nu$ (low $E_\nu$), the $^{10}$Be yield is decreased by factors of $\sim$13-33 (4-10). This significant change shows that the \bepa\ reaction is indeed  a major destruction channel of \iso{10}{Be} in CCSNe and challenges the scenario in which \iso{10}{Be} was injected by a nearby Supernova into the ESS. 

\subsection{Implications for Early Solar System}
\begin{table}
\caption{\label{tab:yields}\iso{10}{Be} yields for the minimum and recommended value of the \bepa\ reaction rate and different models for the neutrino spectra. Yields are given in units of $10^{-10}\,\msun$. Scenarios marked with "($\bar{\nu}_x \leftrightarrow \bar{\nu}_e$)" assume a complete swap of the $\bar{\nu}_x$ and $\bar{\nu}_e$ spectra by flavor transformations. }
\begin{ruledtabular}

    \begin{tabular}{llll}
    Model   & Neutrino Spectrum          & Minimum  & Recommended\\
                     &    &  rate      & rate \\\hline
    1 &Low $E_\nu$  & 0.36 & 0.18 \\
    2 &Simulation   & 0.65 & 0.20 \\
    3 &High $E_\nu$ & 0.43 & 0.12 \\
    4 &Simulation ($\bar{\nu}_x \leftrightarrow \bar{\nu}_e$)  & 0.80 & 0.24 \\
    5 &High $E_\nu$ ($\bar{\nu}_x \leftrightarrow \bar{\nu}_e$)  & 1.23 & 0.34 \\
    \end{tabular}
\end{ruledtabular}
\end{table}
A major problem in reconciling theoretical CCSN nucleosynthesis yields with the abundances of SLRs in the ESS 
is the absence of anomalies in the isotopic composition of common stable isotopes. This excludes most CCSNe \cite{Wasserburg.Busso.ea:2006}, except for those from low-mass progenitors with low explosion energies as the origin of the SLRs \cite{Banerjee.Qian.ea:2016}. 
\citet{Banerjee.Qian.ea:2016} have further demonstrated  that a 11.8~\msun progenitor produces a sufficient  yield  of \iso{10}{Be}  to  simultaneously  explain  the \iso{10}{Be}/\iso{9}{Be}, \iso{41}{Ca}/\iso{40}{Ca} and \iso{107}{Pd}/\iso{108}{Pd} inferred isotopic  ratios, identifying this model as a favorable candidate to explain the ESS SLR abundances. Therefore, we focus in the following on this model and show that, with the new reaction rate discussed above, the yield of the 11.8~\msun model is insufficient to explain the \iso{10}{Be}/\iso{9}{Be} ESS ratio in the scenario proposed by \citet{Banerjee.Qian.ea:2016}. To estimate the required yields, the ESS ratio implied by a CCSN yield $Y_{SLR}$ of a radioactive isotope with mass number $A_{SLR}$ is estimated as
\begin{equation}
    \label{eq:ess_ratio}
    \left( \frac{N_{SLR}}{N_I} \right)_{\mathrm{ESS}} 
    =\frac{f\,Y_{SLR}/A_{SLR}}{X_{I,\odot}M_\odot/A_I}\mathrm{exp}\left( -\Delta/\tau_{SLR} \right),
\end{equation}
where $\tau_{SLR}$ is the decay timescale of the radioactive isotope and we use  $\tau_{SLR}=2.003\,\rm{Myr}$ \cite{Chmeleff:2010} for \iso{10}{Be}. $X_{I,\odot}=1.4\times 10^{-10}$\cite{Asplund.Grevesse.ea:2009} is the solar mass fraction of the stable reference isotope, \iso{9}{Be}, with mass number $A_I$.
There are two free parameters: $f$ is the dilution factor (i.e. fraction of the CCSN ejecta incorporated into the proto-solar cloud) and $\Delta$ is the interval between the production of the SLR by the CCSN and its 
incorporation into ESS solids.
We take a dilution factor of $f=5\times 10^{-4}$ and $\Delta=1$~Myr, which allows us to match the measured ESS ratios of \iso{41}{Ca}/\iso{40}{Ca} and \iso{107}{Pd}/\iso{108}{Pd} for the 11.8~\msun\ model, which is also consistent with constraints from CCSN remnant evolution models and the requirements for the injection of material from the shock into the proto-solar cloud \cite{Banerjee.Qian.ea:2016}. Note that neither \iso{41}{Ca} nor \iso{107}{Pd} are significantly affected by the neutrino spectra or the reaction rates discussed here.
With these parameters, the observed \iso{10}{Be}/\iso{9}{Be} ESS ratio, $(3-9)\times 10^{-4}$ \cite{Liu.Chaussidon.ea:2012} requires yields in the range
$(1.5-4.6)\times 10^{-10}\,\msun$.

\begin{figure}
    \centering
    \includegraphics[width=\linewidth]{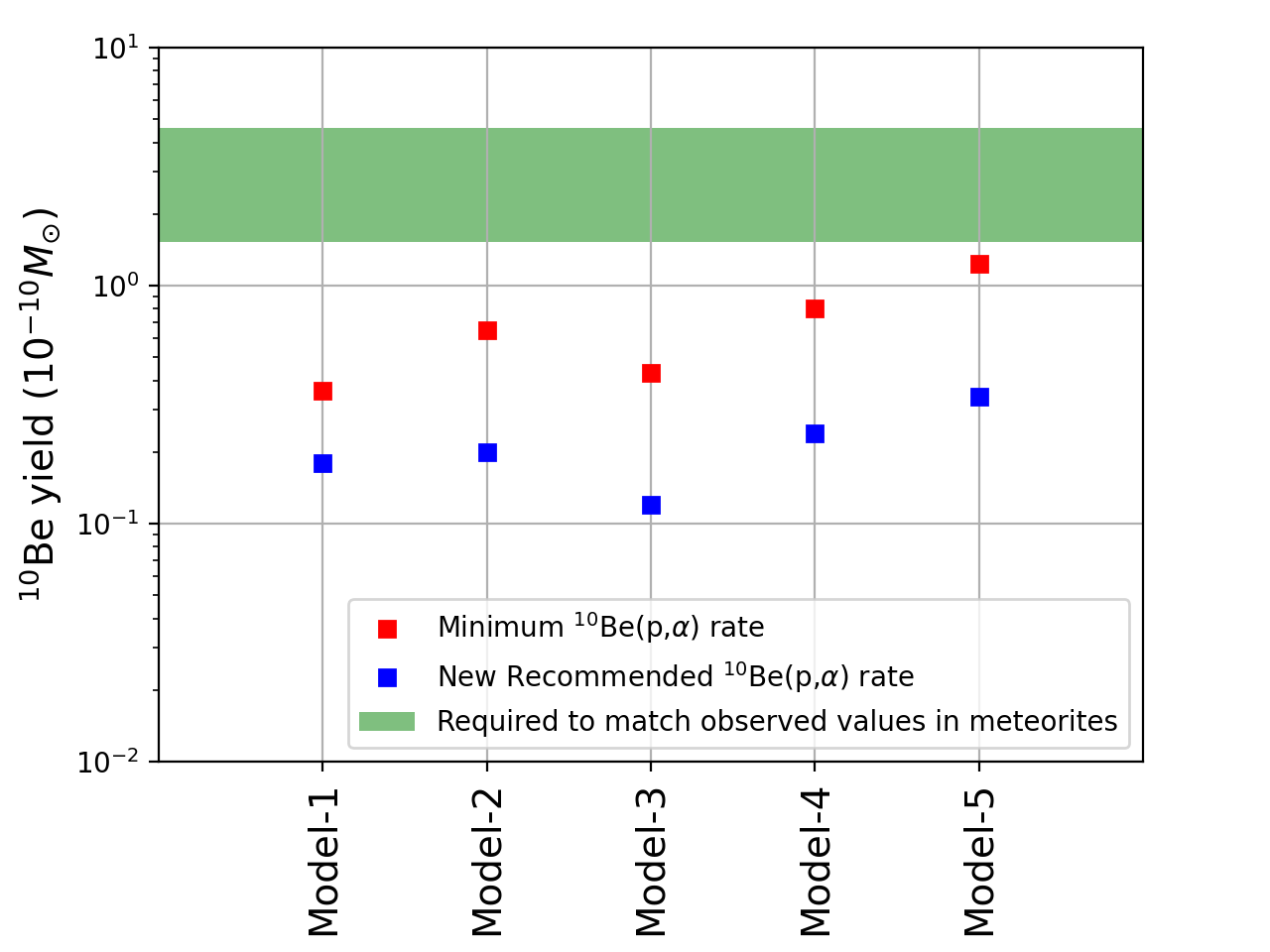}
    \caption{\iso{10}{Be} yield for the different models of the neutrino spectra listed in table \ref{tab:yields} for the recommended and minimum \bepa\ reaction rate. The green band indicates the yield that is required to be in agreement with the observed ESS \iso{10}{Be}/\iso{9}{Be} ratio when assuming the scenario of \cite{Banerjee.Qian.ea:2016}. }
    \label{fig:yield_comparison}
\end{figure}

In order to fully cover the range of uncertainties of the neutrino spectra we calculate the \iso{10}{Be} yield for five different models. We took different assumptions about the neutrino spectra, including low $E_\nu$ (Model-1), the spectra from the simulation \cite{Mueller.Tauris.ea:2019} (Model-2), and high $E_\nu$ (Model-3). We also include estimates for the maximum impact of neutrino flavor transformations by  assuming a complete swap of the spectra of $\bar{\nu}_e$ and $\bar{\nu}_x$ for neutrino spectra from simulation (Model-4) as well as for high $E_{\nu}$ (Model-5).  Table \ref{tab:yields} shows the yields of the 11.8~\msun\ model for the minimum and recommended \bepa\ reaction rate and different neutrino spectra models.  These yield values are also shown in Figure \ref{fig:yield_comparison}, together with band representing the required \iso{10}{Be} yields to match the inferred \iso{10}Be/\iso{9}{Be} ratio. For both values of the \bepa\ reaction rate, the highest \iso{10}{Be} yields result from the high $E_{\nu}$ case with flavor transformations (Model-5), whereas the lowest yields result from the low $E_{\nu}$ case (Model-1).  This shows that all the models, even with the most optimistic neutrino spectra and the minimum \bepa\ reaction rate, fall short of the required \iso{10}{Be} yield in this scenario. 
These findings seem to favor non-thermal processes or in-situ production as main contributors for \iso{10}{Be} in the ESS.  

\section{\label{sec:conclusions}Conclusions}
We have studied the production of \iso{10}{Be} by the $\nu$ process in CCSN explosions for a range of progenitors and for different models of the neutrino spectra, to investigate the role of the \bepa, \bean\ and  \benue\ reactions as \iso{10}{Be} destruction channels. An enhanced \bepa\ reaction rate significantly reduces the yields of this SLR and it changes the relationship between the neutrino spectra and the net \iso{10}{Be} yield. The large uncertainty of the \bean\ reaction rate can change the \iso{10}{Be} yield by a factor of~2, making it the second most important reaction, after the \bepa\ reaction, as a destruction channel for \iso{10}{Be} in CCSNe.

We re-evaluated the \bepa\ reaction rate using $R$-matrix calculations based on updated nuclear data including the recently observed low energy resonance at 193~\kev\ in the compound nucleus \iso{11}B \cite{Ayyad19}. We show that this resonance makes the \bepa\ reaction rate significantly larger (up to a factor of 20) than the rate in the JINA-REACLIB database. With our recommended rate, the \iso{10}{Be} yields are up to a factor of 33 lower than those obtained with the default REACLIB rate. However, further experimental constraints on the proton and alpha widths of the 193 keV resonance are necessary to narrow down the range of calculated \iso{10}{Be} yields. We show that, with current CCSNe models and  the estimated uncertainty of the \bepa\ reaction rate and neutrino spectra, a nearby low-mass CCSN (11.8~\msun\ model) cannot explain the \iso{10}{Be}/\iso{9}{Be} ratio in agreement with other isotopic ratios inferred from meteorites. These findings point towards non-thermal, in-situ production of \iso{10}{Be} in the ESS. We encourage the further refinement in the prediction of CCSN neutrino spectra (including flavor transformations), as well as  experiments to constrain the properties of 193~\kev resonance in the \bepa\ reaction and the \bean\ reaction rate to further constrain our findings.

\acknowledgements This work was supported by the U.S. Department of
Energy, Office of Science, Office of Nuclear Physics and Office of
Advanced Scientific Computing Research, Scientific Discovery through
Advanced Computing (SciDAC) program. Research at Oak Ridge National
Laboratory is supported under contract DE-AC05-00OR22725 from the
U.S. Department of Energy to UT-Battelle, LLC. JSR, RJD and TA were supported by National Science Foundation through Grant No. Phys-2011890. RJD utilized resources from the Notre Dame Center for Research Computing and was supported by the Joint Institute for Nuclear Astrophysics through Grant
No. PHY-1430152 (JINA Center for the Evolution of the Elements). RM
and GMP acknowledge the support of the Deutsche Forschungsgemeinschaft
(DFG, German Research Foundation) -- Project-ID 279384907 -- SFB 1245
``Nuclei: From Fundamental Interactions to Structure and Stars''. 
This research made extensive use of numpy \cite{numpy1}, matplotlib \cite{Hunter:2007} and of the SAO/NASA Astrophysics Data System (ADS).
\appendix

\begin{figure}
    \centering
    \includegraphics[width=\linewidth]{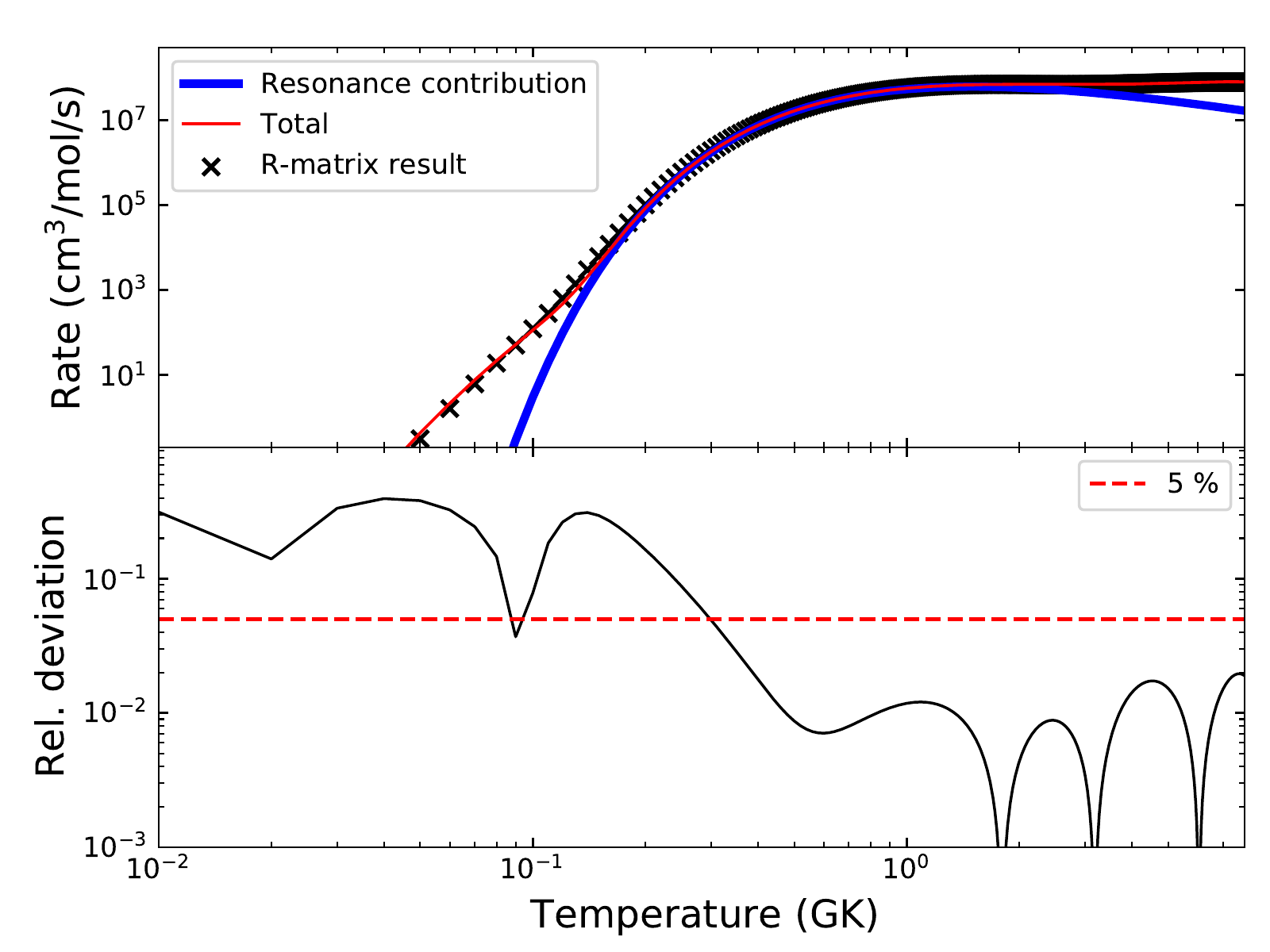}
    \caption{\label{fig:fit}Reaction rate fit with the REACLIB parameterization, using a resonant and non-resonant contribution. The parameterization and reaction rate calculation agree within 5~\% for all temperatures above 0.3 GK, which are the most relevant for CCSN explosions.}
    \label{fig:reaclib_fit}
\end{figure}

\section{\label{sec:reaclib_fit}\texorpdfstring{\bepa}{10Be(p,a)7Li} rate: REACLIB fit parameters}
In the REACLIB reaction rate library, thermonuclear reaction rates
are described as a function of temperature in GK, $T$, by a seven
parameter fit as
\begin{equation}
    \label{eq:fit}
    \langle\sigma v\rangle(T)=\exp\left[
      a_0 + \left(\sum\limits_{i=1}^{5} a_i\, T^{\frac{2i-5}{3}} \right) + a_6\, \rm{ln}(T)
    \right].
\end{equation}

Resonant and non-resonant contributions need to be fitted by separate sets of parameters.
Table \ref{tab:fit_parameters} gives the fit parameters for the minimum and recommended rate for the \bepa\ reaction discussed in \ref{sec:data}. Each fit consists of two sets of seven parameters, for the resonant and non-resonant contributions. The fit is illustrated for the recommended rate in Fig. \ref{fig:fit}. Deviations of the fit from the R-matrix calculation are less than $5\,\%$ for temperatures higher than $0.3\,\rm{GK}$.
The parameters for the reverse reaction rate can be obtained from the forward reaction by detailed balance by replacing $a_0^{\rm{rev}}=a_0-2.2375$ and $a_1^{\rm{rev}}=a_1-29.7539$.
The other parameters $a_{2\dots6}$ remain the same \cite{Rauscher.Thielemann:2000}.
\begin{table}[H]
\caption{Fit parameters for the minimum and recommended rate. Each rate consists of a resonant and non-resonant contribution.}
\begin{ruledtabular}
    \begin{tabular}{l|rr|rr}
         & \multicolumn{2}{c|}{Minimum} &\multicolumn{2}{c}{Recommended} \\ 
        & resonant & non-resonant & resonant & non-resonant \\ \hline
     $a_0$ & 18.83813& 30.49055    &  20.01675  & 29.05572    \\
     $a_1$ &-2.236187& 0.0         &   -2.236187  &  0.0   \\
     $a_2$ &0.0      & -11.32177   &   0.0 & -11.25624    \\
     $a_3$ &0.0      & -9.265300   &   0.0 &   -3.687460  \\
     $a_4$ &0.0      & 3.559158    &   0.0 &  0.7607396  \\
     $a_5$ &0.0      & -0.5154761  &   0.0 &   0.08781213 \\
     $a_6$ &-1.5     &-$2/3$       & -1.5   & -$2/3$    \\
    \end{tabular}
    \label{tab:fit_parameters}
\end{ruledtabular} 
\end{table}

\bibliography{bib}

\begin{thebibliography}{55}%
\makeatletter
\providecommand \@ifxundefined [1]{%
 \@ifx{#1\undefined}
}%
\providecommand \@ifnum [1]{%
 \ifnum #1\expandafter \@firstoftwo
 \else \expandafter \@secondoftwo
 \fi
}%
\providecommand \@ifx [1]{%
 \ifx #1\expandafter \@firstoftwo
 \else \expandafter \@secondoftwo
 \fi
}%
\providecommand \natexlab [1]{#1}%
\providecommand \enquote  [1]{``#1''}%
\providecommand \bibnamefont  [1]{#1}%
\providecommand \bibfnamefont [1]{#1}%
\providecommand \citenamefont [1]{#1}%
\providecommand \href@noop [0]{\@secondoftwo}%
\providecommand \href [0]{\begingroup \@sanitize@url \@href}%
\providecommand \@href[1]{\@@startlink{#1}\@@href}%
\providecommand \@@href[1]{\endgroup#1\@@endlink}%
\providecommand \@sanitize@url [0]{\catcode `\\12\catcode `\$12\catcode
  `\&12\catcode `\#12\catcode `\^12\catcode `\_12\catcode `\%12\relax}%
\providecommand \@@startlink[1]{}%
\providecommand \@@endlink[0]{}%
\providecommand \url  [0]{\begingroup\@sanitize@url \@url }%
\providecommand \@url [1]{\endgroup\@href {#1}{\urlprefix }}%
\providecommand \urlprefix  [0]{URL }%
\providecommand \Eprint [0]{\href }%
\providecommand \doibase [0]{https://doi.org/}%
\providecommand \selectlanguage [0]{\@gobble}%
\providecommand \bibinfo  [0]{\@secondoftwo}%
\providecommand \bibfield  [0]{\@secondoftwo}%
\providecommand \translation [1]{[#1]}%
\providecommand \BibitemOpen [0]{}%
\providecommand \bibitemStop [0]{}%
\providecommand \bibitemNoStop [0]{.\EOS\space}%
\providecommand \EOS [0]{\spacefactor3000\relax}%
\providecommand \BibitemShut  [1]{\csname bibitem#1\endcsname}%
\let\auto@bib@innerbib\@empty
\bibitem [{\citenamefont {{Lee}}\ \emph {et~al.}(1977)\citenamefont {{Lee}},
  \citenamefont {{Papanastassiou}},\ and\ \citenamefont
  {{Wasserburg}}}]{Lee.Papanastassiou.ea:1977}%
  \BibitemOpen
  \bibfield  {author} {\bibinfo {author} {\bibfnamefont {T.}~\bibnamefont
  {{Lee}}}, \bibinfo {author} {\bibfnamefont {D.~A.}\ \bibnamefont
  {{Papanastassiou}}},\ and\ \bibinfo {author} {\bibfnamefont {G.~J.}\
  \bibnamefont {{Wasserburg}}},\ }\bibfield  {title} {\bibinfo {title}
  {{Aluminum-26 in the early solar system: fossil or fuel?}},\ }\href
  {https://doi.org/10.1086/182351} {\bibfield  {journal} {\bibinfo  {journal}
  {Astrophys. J. Lett.}\ }\textbf {\bibinfo {volume} {211}},\ \bibinfo {pages}
  {L107} (\bibinfo {year} {1977})}\BibitemShut {NoStop}%
\bibitem [{\citenamefont {{Reynolds}}(1960)}]{Reynolds:1960}%
  \BibitemOpen
  \bibfield  {author} {\bibinfo {author} {\bibfnamefont {J.~H.}\ \bibnamefont
  {{Reynolds}}},\ }\bibfield  {title} {\bibinfo {title} {{Determination of the
  Age of the Elements}},\ }\href {https://doi.org/10.1103/PhysRevLett.4.8}
  {\bibfield  {journal} {\bibinfo  {journal} {Phys. Rev. Lett.}\ }\textbf
  {\bibinfo {volume} {4}},\ \bibinfo {pages} {8} (\bibinfo {year}
  {1960})}\BibitemShut {NoStop}%
\bibitem [{\citenamefont {{Dauphas}}\ and\ \citenamefont
  {{Chaussidon}}(2011)}]{Dauphas.Chaussidon:2011}%
  \BibitemOpen
  \bibfield  {author} {\bibinfo {author} {\bibfnamefont {N.}~\bibnamefont
  {{Dauphas}}}\ and\ \bibinfo {author} {\bibfnamefont {M.}~\bibnamefont
  {{Chaussidon}}},\ }\bibfield  {title} {\bibinfo {title} {{A Perspective from
  Extinct Radionuclides on a Young Stellar Object: The Sun and Its Accretion
  Disk}},\ }\href {https://doi.org/10.1146/annurev-earth-040610-133428}
  {\bibfield  {journal} {\bibinfo  {journal} {Annu. Rev. of Earth and Planet.
  Sci.}\ }\textbf {\bibinfo {volume} {39}},\ \bibinfo {pages} {351} (\bibinfo
  {year} {2011})}\BibitemShut {NoStop}%
\bibitem [{\citenamefont {{Lugaro}}\ \emph {et~al.}(2018)\citenamefont
  {{Lugaro}}, \citenamefont {{Ott}},\ and\ \citenamefont
  {{Kereszturi}}}]{Lugaro.Ott.ea:2018}%
  \BibitemOpen
  \bibfield  {author} {\bibinfo {author} {\bibfnamefont {M.}~\bibnamefont
  {{Lugaro}}}, \bibinfo {author} {\bibfnamefont {U.}~\bibnamefont {{Ott}}},\
  and\ \bibinfo {author} {\bibfnamefont {{\'A}.}~\bibnamefont {{Kereszturi}}},\
  }\bibfield  {title} {\bibinfo {title} {{Radioactive nuclei from
  cosmochronology to habitability}},\ }\href
  {https://doi.org/10.1016/j.ppnp.2018.05.002} {\bibfield  {journal} {\bibinfo
  {journal} {Prog. Part. Nucl. Phys.}\ }\textbf {\bibinfo {volume} {102}},\
  \bibinfo {pages} {1} (\bibinfo {year} {2018})}\BibitemShut {NoStop}%
\bibitem [{\citenamefont {{Zucker}}\ \emph {et~al.}(2022)\citenamefont
  {{Zucker}}, \citenamefont {{Goodman}}, \citenamefont {{Alves}}, \citenamefont
  {{Bialy}}, \citenamefont {{Foley}}, \citenamefont {{Speagle}}, \citenamefont
  {{Gro\ss{}schedl}}, \citenamefont {{Finkbeiner}}, \citenamefont {{Burkert}},
  \citenamefont {{Khimey}},\ and\ \citenamefont
  {{Swiggum}}}]{Zucker.Goodman.ea:2022}%
  \BibitemOpen
  \bibfield  {author} {\bibinfo {author} {\bibfnamefont {C.}~\bibnamefont
  {{Zucker}}}, \bibinfo {author} {\bibfnamefont {A.~A.}\ \bibnamefont
  {{Goodman}}}, \bibinfo {author} {\bibfnamefont {J.}~\bibnamefont {{Alves}}},
  \bibinfo {author} {\bibfnamefont {S.}~\bibnamefont {{Bialy}}}, \bibinfo
  {author} {\bibfnamefont {M.}~\bibnamefont {{Foley}}}, \bibinfo {author}
  {\bibfnamefont {J.~S.}\ \bibnamefont {{Speagle}}}, \bibinfo {author}
  {\bibfnamefont {J.}~\bibnamefont {{Gro\ss{}schedl}}}, \bibinfo {author}
  {\bibfnamefont {D.~P.}\ \bibnamefont {{Finkbeiner}}}, \bibinfo {author}
  {\bibfnamefont {A.}~\bibnamefont {{Burkert}}}, \bibinfo {author}
  {\bibfnamefont {D.}~\bibnamefont {{Khimey}}},\ and\ \bibinfo {author}
  {\bibfnamefont {C.}~\bibnamefont {{Swiggum}}},\ }\bibfield  {title} {\bibinfo
  {title} {{Star formation near the Sun is driven by expansion of the Local
  Bubble}},\ }\href {https://doi.org/10.1038/s41586-021-04286-5} {\bibfield
  {journal} {\bibinfo  {journal} {Nature}\ }\textbf {\bibinfo {volume} {601}},\
  \bibinfo {pages} {334} (\bibinfo {year} {2022})}\BibitemShut {NoStop}%
\bibitem [{\citenamefont {{Forbes}}\ \emph {et~al.}(2021)\citenamefont
  {{Forbes}}, \citenamefont {{Alves}},\ and\ \citenamefont
  {{Lin}}}]{Forbes.Alves.ea:2021}%
  \BibitemOpen
  \bibfield  {author} {\bibinfo {author} {\bibfnamefont {J.~C.}\ \bibnamefont
  {{Forbes}}}, \bibinfo {author} {\bibfnamefont {J.}~\bibnamefont {{Alves}}},\
  and\ \bibinfo {author} {\bibfnamefont {D.~N.~C.}\ \bibnamefont {{Lin}}},\
  }\bibfield  {title} {\bibinfo {title} {{A Solar System formation analogue in
  the Ophiuchus star-forming complex}},\ }\href
  {https://doi.org/10.1038/s41550-021-01442-9} {\bibfield  {journal} {\bibinfo
  {journal} {Nature Astronomy}\ }\textbf {\bibinfo {volume} {5}},\ \bibinfo
  {pages} {1009} (\bibinfo {year} {2021})}\BibitemShut {NoStop}%
\bibitem [{\citenamefont {{Krause}}\ \emph {et~al.}(2018)\citenamefont
  {{Krause}}, \citenamefont {{Burkert}}, \citenamefont {{Diehl}}, \citenamefont
  {{Fierlinger}}, \citenamefont {{Gaczkowski}}, \citenamefont {{Kroell}},
  \citenamefont {{Ngoumou}}, \citenamefont {{Roccatagliata}}, \citenamefont
  {{Siegert}},\ and\ \citenamefont {{Preibisch}}}]{Krause.Burkert.ea:2018}%
  \BibitemOpen
  \bibfield  {author} {\bibinfo {author} {\bibfnamefont {M.~G.~H.}\
  \bibnamefont {{Krause}}}, \bibinfo {author} {\bibfnamefont {A.}~\bibnamefont
  {{Burkert}}}, \bibinfo {author} {\bibfnamefont {R.}~\bibnamefont {{Diehl}}},
  \bibinfo {author} {\bibfnamefont {K.}~\bibnamefont {{Fierlinger}}}, \bibinfo
  {author} {\bibfnamefont {B.}~\bibnamefont {{Gaczkowski}}}, \bibinfo {author}
  {\bibfnamefont {D.}~\bibnamefont {{Kroell}}}, \bibinfo {author}
  {\bibfnamefont {J.}~\bibnamefont {{Ngoumou}}}, \bibinfo {author}
  {\bibfnamefont {V.}~\bibnamefont {{Roccatagliata}}}, \bibinfo {author}
  {\bibfnamefont {T.}~\bibnamefont {{Siegert}}},\ and\ \bibinfo {author}
  {\bibfnamefont {T.}~\bibnamefont {{Preibisch}}},\ }\bibfield  {title}
  {\bibinfo {title} {{Surround and Squash: the impact of superbubbles on the
  interstellar medium in Scorpius-Centaurus OB2}},\ }\href
  {https://doi.org/10.1051/0004-6361/201732416} {\bibfield  {journal} {\bibinfo
   {journal} {Astron. Astrophys.}\ }\textbf {\bibinfo {volume} {619}},\
  \bibinfo {eid} {A120} (\bibinfo {year} {2018})}\BibitemShut {NoStop}%
\bibitem [{\citenamefont {Cameron}\ and\ \citenamefont
  {Truran}(1977)}]{Cameron.Truran:1977}%
  \BibitemOpen
  \bibfield  {author} {\bibinfo {author} {\bibfnamefont {A.}~\bibnamefont
  {Cameron}}\ and\ \bibinfo {author} {\bibfnamefont {J.}~\bibnamefont
  {Truran}},\ }\bibfield  {title} {\bibinfo {title} {The supernova trigger for
  formation of the solar system},\ }\href
  {https://doi.org/https://doi.org/10.1016/0019-1035(77)90101-4} {\bibfield
  {journal} {\bibinfo  {journal} {Icarus}\ }\textbf {\bibinfo {volume} {30}},\
  \bibinfo {pages} {447 } (\bibinfo {year} {1977})}\BibitemShut {NoStop}%
\bibitem [{\citenamefont {{Chmeleff}}\ \emph {et~al.}(2010)\citenamefont
  {{Chmeleff}}, \citenamefont {{von Blanckenburg}}, \citenamefont {{Kossert}},\
  and\ \citenamefont {{Jakob}}}]{Chmeleff:2010}%
  \BibitemOpen
  \bibfield  {author} {\bibinfo {author} {\bibfnamefont {J.}~\bibnamefont
  {{Chmeleff}}}, \bibinfo {author} {\bibfnamefont {F.}~\bibnamefont {{von
  Blanckenburg}}}, \bibinfo {author} {\bibfnamefont {K.}~\bibnamefont
  {{Kossert}}},\ and\ \bibinfo {author} {\bibfnamefont {D.}~\bibnamefont
  {{Jakob}}},\ }\bibfield  {title} {\bibinfo {title} {{Determination of the
  $^{10}$Be half-life by multicollector ICP-MS and liquid scintillation
  counting}},\ }\href {https://doi.org/10.1016/j.nimb.2009.09.012} {\bibfield
  {journal} {\bibinfo  {journal} {Nuclear Instruments and Methods in Physics
  Research B}\ }\textbf {\bibinfo {volume} {268}},\ \bibinfo {pages} {192}
  (\bibinfo {year} {2010})}\BibitemShut {NoStop}%
\bibitem [{\citenamefont {{Fukuda}}\ \emph {et~al.}(2019)\citenamefont
  {{Fukuda}}, \citenamefont {{Hiyagon}}, \citenamefont {{Fujiya}},
  \citenamefont {{Takahata}}, \citenamefont {{Kagoshima}},\ and\ \citenamefont
  {{Sano}}}]{Fukuda.Hiyagon.ea:2019}%
  \BibitemOpen
  \bibfield  {author} {\bibinfo {author} {\bibfnamefont {K.}~\bibnamefont
  {{Fukuda}}}, \bibinfo {author} {\bibfnamefont {H.}~\bibnamefont {{Hiyagon}}},
  \bibinfo {author} {\bibfnamefont {W.}~\bibnamefont {{Fujiya}}}, \bibinfo
  {author} {\bibfnamefont {N.}~\bibnamefont {{Takahata}}}, \bibinfo {author}
  {\bibfnamefont {T.}~\bibnamefont {{Kagoshima}}},\ and\ \bibinfo {author}
  {\bibfnamefont {Y.}~\bibnamefont {{Sano}}},\ }\bibfield  {title} {\bibinfo
  {title} {{Origin of the Short-lived Radionuclide $^{10}$Be and Its
  Implications for the Astronomical Setting of CAI Formation in the Solar
  Protoplanetary Disk}},\ }\href {https://doi.org/10.3847/1538-4357/ab479c}
  {\bibfield  {journal} {\bibinfo  {journal} {Astrophys. J.}\ }\textbf
  {\bibinfo {volume} {886}},\ \bibinfo {eid} {34} (\bibinfo {year}
  {2019})}\BibitemShut {NoStop}%
\bibitem [{\citenamefont {{Tatischeff}}\ \emph {et~al.}(2014)\citenamefont
  {{Tatischeff}}, \citenamefont {{Duprat}},\ and\ \citenamefont {{de
  S{\'e}r{\'e}ville}}}]{Tatischeff.Duprat.ea:2014}%
  \BibitemOpen
  \bibfield  {author} {\bibinfo {author} {\bibfnamefont {V.}~\bibnamefont
  {{Tatischeff}}}, \bibinfo {author} {\bibfnamefont {J.}~\bibnamefont
  {{Duprat}}},\ and\ \bibinfo {author} {\bibfnamefont {N.}~\bibnamefont {{de
  S{\'e}r{\'e}ville}}},\ }\bibfield  {title} {\bibinfo {title} {{Light-element
  Nucleosynthesis in a Molecular Cloud Interacting with a Supernova Remnant and
  the Origin of Beryllium-10 in the Protosolar Nebula}},\ }\href
  {https://doi.org/10.1088/0004-637X/796/2/124} {\bibfield  {journal} {\bibinfo
   {journal} {Astrophys. J.}\ }\textbf {\bibinfo {volume} {796}},\ \bibinfo
  {eid} {124} (\bibinfo {year} {2014})}\BibitemShut {NoStop}%
\bibitem [{\citenamefont {{Duprat}}\ and\ \citenamefont
  {{Tatischeff}}(2007)}]{Duprat.Tatischeff.ea:2007}%
  \BibitemOpen
  \bibfield  {author} {\bibinfo {author} {\bibfnamefont {J.}~\bibnamefont
  {{Duprat}}}\ and\ \bibinfo {author} {\bibfnamefont {V.}~\bibnamefont
  {{Tatischeff}}},\ }\bibfield  {title} {\bibinfo {title} {{Energetic
  Constraints on In Situ Production of Short-Lived Radionuclei in the Early
  Solar System}},\ }\href {https://doi.org/10.1086/524297} {\bibfield
  {journal} {\bibinfo  {journal} {Astrophys. J. Lett.}\ }\textbf {\bibinfo
  {volume} {671}},\ \bibinfo {pages} {L69} (\bibinfo {year}
  {2007})}\BibitemShut {NoStop}%
\bibitem [{\citenamefont {{Desch}}\ \emph {et~al.}(2004)\citenamefont
  {{Desch}}, \citenamefont {{Connolly}},\ and\ \citenamefont
  {{Srinivasan}}}]{Desch.Connolly.ea:2004}%
  \BibitemOpen
  \bibfield  {author} {\bibinfo {author} {\bibfnamefont {S.~J.}\ \bibnamefont
  {{Desch}}}, \bibinfo {author} {\bibfnamefont {J.}~\bibnamefont {{Connolly}},
  \bibfnamefont {Harold~C.}},\ and\ \bibinfo {author} {\bibfnamefont
  {G.}~\bibnamefont {{Srinivasan}}},\ }\bibfield  {title} {\bibinfo {title}
  {{An Interstellar Origin for the Beryllium 10 in Calcium-rich, Aluminum-rich
  Inclusions}},\ }\href {https://doi.org/10.1086/380831} {\bibfield  {journal}
  {\bibinfo  {journal} {Astrophys. J.}\ }\textbf {\bibinfo {volume} {602}},\
  \bibinfo {pages} {528} (\bibinfo {year} {2004})}\BibitemShut {NoStop}%
\bibitem [{\citenamefont {{Lee}}\ \emph {et~al.}(1998)\citenamefont {{Lee}},
  \citenamefont {{Shu}}, \citenamefont {{Shang}}, \citenamefont {{Glassgold}},\
  and\ \citenamefont {{Rehm}}}]{Lee.Shu.ea:1998}%
  \BibitemOpen
  \bibfield  {author} {\bibinfo {author} {\bibfnamefont {T.}~\bibnamefont
  {{Lee}}}, \bibinfo {author} {\bibfnamefont {F.~H.}\ \bibnamefont {{Shu}}},
  \bibinfo {author} {\bibfnamefont {H.}~\bibnamefont {{Shang}}}, \bibinfo
  {author} {\bibfnamefont {A.~E.}\ \bibnamefont {{Glassgold}}},\ and\ \bibinfo
  {author} {\bibfnamefont {K.~E.}\ \bibnamefont {{Rehm}}},\ }\bibfield  {title}
  {\bibinfo {title} {{Protostellar Cosmic Rays and Extinct Radioactivities in
  Meteorites}},\ }\href {https://doi.org/10.1086/306284} {\bibfield  {journal}
  {\bibinfo  {journal} {Astrophys. J.}\ }\textbf {\bibinfo {volume} {506}},\
  \bibinfo {pages} {898} (\bibinfo {year} {1998})}\BibitemShut {NoStop}%
\bibitem [{\citenamefont {{Woosley}}\ \emph {et~al.}(1990)\citenamefont
  {{Woosley}}, \citenamefont {{Hartmann}}, \citenamefont {{Hoffman}},\ and\
  \citenamefont {{Haxton}}}]{Woosley.Hartmann.ea:1990}%
  \BibitemOpen
  \bibfield  {author} {\bibinfo {author} {\bibfnamefont {S.~E.}\ \bibnamefont
  {{Woosley}}}, \bibinfo {author} {\bibfnamefont {D.~H.}\ \bibnamefont
  {{Hartmann}}}, \bibinfo {author} {\bibfnamefont {R.~D.}\ \bibnamefont
  {{Hoffman}}},\ and\ \bibinfo {author} {\bibfnamefont {W.~C.}\ \bibnamefont
  {{Haxton}}},\ }\bibfield  {title} {\bibinfo {title} {The $\nu$-process},\
  }\href {https://doi.org/10.1086/168839} {\bibfield  {journal} {\bibinfo
  {journal} {Astrophys. J.}\ }\textbf {\bibinfo {volume} {356}},\ \bibinfo
  {pages} {272} (\bibinfo {year} {1990})}\BibitemShut {NoStop}%
\bibitem [{\citenamefont {{Balasi}}\ \emph {et~al.}(2015)\citenamefont
  {{Balasi}}, \citenamefont {{Langanke}},\ and\ \citenamefont
  {{Mart{\'\i}nez-Pinedo}}}]{Balasi.Langanke.ea:2015}%
  \BibitemOpen
  \bibfield  {author} {\bibinfo {author} {\bibfnamefont {K.~G.}\ \bibnamefont
  {{Balasi}}}, \bibinfo {author} {\bibfnamefont {K.}~\bibnamefont
  {{Langanke}}},\ and\ \bibinfo {author} {\bibfnamefont {G.}~\bibnamefont
  {{Mart{\'\i}nez-Pinedo}}},\ }\bibfield  {title} {\bibinfo {title}
  {{Neutrino-nucleus reactions and their role for supernova dynamics and
  nucleosynthesis}},\ }\href {https://doi.org/10.1016/j.ppnp.2015.08.001}
  {\bibfield  {journal} {\bibinfo  {journal} {Progress in Particle and Nuclear
  Physics}\ }\textbf {\bibinfo {volume} {85}},\ \bibinfo {pages} {33} (\bibinfo
  {year} {2015})}\BibitemShut {NoStop}%
\bibitem [{\citenamefont {{Banerjee}}\ \emph {et~al.}(2016)\citenamefont
  {{Banerjee}}, \citenamefont {{Qian}}, \citenamefont {{Heger}},\ and\
  \citenamefont {{Haxton}}}]{Banerjee.Qian.ea:2016}%
  \BibitemOpen
  \bibfield  {author} {\bibinfo {author} {\bibfnamefont {P.}~\bibnamefont
  {{Banerjee}}}, \bibinfo {author} {\bibfnamefont {Y.-Z.}\ \bibnamefont
  {{Qian}}}, \bibinfo {author} {\bibfnamefont {A.}~\bibnamefont {{Heger}}},\
  and\ \bibinfo {author} {\bibfnamefont {W.~C.}\ \bibnamefont {{Haxton}}},\
  }\bibfield  {title} {\bibinfo {title} {{Evidence from stable isotopes and
  $^{10}$Be for solar system formation triggered by a low-mass supernova}},\
  }\href {https://doi.org/10.1038/ncomms13639} {\bibfield  {journal} {\bibinfo
  {journal} {Nature Commun.}\ }\textbf {\bibinfo {volume} {7}},\ \bibinfo {eid}
  {13639} (\bibinfo {year} {2016})}\BibitemShut {NoStop}%
\bibitem [{\citenamefont {{Tang}}\ and\ \citenamefont
  {{Dauphas}}(2015)}]{Tang.Dauphas:2015}%
  \BibitemOpen
  \bibfield  {author} {\bibinfo {author} {\bibfnamefont {H.}~\bibnamefont
  {{Tang}}}\ and\ \bibinfo {author} {\bibfnamefont {N.}~\bibnamefont
  {{Dauphas}}},\ }\bibfield  {title} {\bibinfo {title} {{Low $^{60}$Fe
  Abundance in Semarkona and Sahara 99555}},\ }\href
  {https://doi.org/10.1088/0004-637X/802/1/22} {\bibfield  {journal} {\bibinfo
  {journal} {Astrophys. J.}\ }\textbf {\bibinfo {volume} {802}},\ \bibinfo
  {eid} {22} (\bibinfo {year} {2015})}\BibitemShut {NoStop}%
\bibitem [{\citenamefont {{Tang}}\ and\ \citenamefont
  {{Dauphas}}(2012)}]{Tang.Dauphas.ea:2012}%
  \BibitemOpen
  \bibfield  {author} {\bibinfo {author} {\bibfnamefont {H.}~\bibnamefont
  {{Tang}}}\ and\ \bibinfo {author} {\bibfnamefont {N.}~\bibnamefont
  {{Dauphas}}},\ }\bibfield  {title} {\bibinfo {title} {{Abundance,
  distribution, and origin of $^{60}$Fe in the solar protoplanetary disk}},\
  }\href {https://doi.org/10.1016/j.epsl.2012.10.011} {\bibfield  {journal}
  {\bibinfo  {journal} {Earth and Planet. Sci. Lett.}\ }\textbf {\bibinfo
  {volume} {359}},\ \bibinfo {pages} {248} (\bibinfo {year}
  {2012})}\BibitemShut {NoStop}%
\bibitem [{\citenamefont {{Cyburt}}\ \emph {et~al.}(2010)\citenamefont
  {{Cyburt}}, \citenamefont {{Amthor}}, \citenamefont {{Ferguson}},
  \citenamefont {{Meisel}}, \citenamefont {{Smith}}, \citenamefont {{Warren}},
  \citenamefont {{Heger}}, \citenamefont {{Hoffman}}, \citenamefont
  {{Rauscher}}, \citenamefont {{Sakharuk}}, \citenamefont {{Schatz}},
  \citenamefont {{Thielemann}},\ and\ \citenamefont
  {{Wiescher}}}]{Cyburt.Amthor.ea:2010}%
  \BibitemOpen
  \bibfield  {author} {\bibinfo {author} {\bibfnamefont {R.~H.}\ \bibnamefont
  {{Cyburt}}}, \bibinfo {author} {\bibfnamefont {A.~M.}\ \bibnamefont
  {{Amthor}}}, \bibinfo {author} {\bibfnamefont {R.}~\bibnamefont
  {{Ferguson}}}, \bibinfo {author} {\bibfnamefont {Z.}~\bibnamefont
  {{Meisel}}}, \bibinfo {author} {\bibfnamefont {K.}~\bibnamefont {{Smith}}},
  \bibinfo {author} {\bibfnamefont {S.}~\bibnamefont {{Warren}}}, \bibinfo
  {author} {\bibfnamefont {A.}~\bibnamefont {{Heger}}}, \bibinfo {author}
  {\bibfnamefont {R.~D.}\ \bibnamefont {{Hoffman}}}, \bibinfo {author}
  {\bibfnamefont {T.}~\bibnamefont {{Rauscher}}}, \bibinfo {author}
  {\bibfnamefont {A.}~\bibnamefont {{Sakharuk}}}, \bibinfo {author}
  {\bibfnamefont {H.}~\bibnamefont {{Schatz}}}, \bibinfo {author}
  {\bibfnamefont {F.~K.}\ \bibnamefont {{Thielemann}}},\ and\ \bibinfo {author}
  {\bibfnamefont {M.}~\bibnamefont {{Wiescher}}},\ }\bibfield  {title}
  {\bibinfo {title} {{The JINA REACLIB Database: Its Recent Updates and Impact
  on Type-I X-ray Bursts}},\ }\href
  {https://doi.org/10.1088/0067-0049/189/1/240} {\bibfield  {journal} {\bibinfo
   {journal} {Astrophys. J. Suppl. Ser.}\ }\textbf {\bibinfo {volume} {189}},\
  \bibinfo {pages} {240} (\bibinfo {year} {2010})}\BibitemShut {NoStop}%
\bibitem [{Note1()}]{Note1}%
  \BibitemOpen
  \bibinfo {note} {\protect \url
  {https://reaclib.jinaweb.org/index.php}}\BibitemShut {NoStop}%
\bibitem [{\citenamefont {{Wagoner}}(1969)}]{Wagoner:1969}%
  \BibitemOpen
  \bibfield  {author} {\bibinfo {author} {\bibfnamefont {R.~V.}\ \bibnamefont
  {{Wagoner}}},\ }\bibfield  {title} {\bibinfo {title} {{Synthesis of the
  Elements Within Objects Exploding from Very High Temperatures}},\ }\href
  {https://doi.org/10.1086/190191} {\bibfield  {journal} {\bibinfo  {journal}
  {Astrophys. J. Suppl. Ser.}\ }\textbf {\bibinfo {volume} {18}},\ \bibinfo
  {pages} {247} (\bibinfo {year} {1969})}\BibitemShut {NoStop}%
\bibitem [{\citenamefont {{Sieverding}}\ \emph {et~al.}(2020)\citenamefont
  {{Sieverding}}, \citenamefont {{M{\"u}ller}},\ and\ \citenamefont
  {{Qian}}}]{Sieverding.Mueller.ea:2021}%
  \BibitemOpen
  \bibfield  {author} {\bibinfo {author} {\bibfnamefont {A.}~\bibnamefont
  {{Sieverding}}}, \bibinfo {author} {\bibfnamefont {B.}~\bibnamefont
  {{M{\"u}ller}}},\ and\ \bibinfo {author} {\bibfnamefont {Y.~Z.}\ \bibnamefont
  {{Qian}}},\ }\bibfield  {title} {\bibinfo {title} {{Nucleosynthesis of an
  11.8 M$_{{\ensuremath{\odot}}}$ Supernova with 3D Simulation of the Inner
  Ejecta: Overall Yields and Implications for Short-lived Radionuclides in the
  Early Solar System}},\ }\href {https://doi.org/10.3847/1538-4357/abc61b}
  {\bibfield  {journal} {\bibinfo  {journal} {Astrophys. J.}\ }\textbf
  {\bibinfo {volume} {904}},\ \bibinfo {eid} {163} (\bibinfo {year}
  {2020})}\BibitemShut {NoStop}%
\bibitem [{\citenamefont {{Sieverding}}\ \emph {et~al.}(2018)\citenamefont
  {{Sieverding}}, \citenamefont {{Mart{\'\i}nez-Pinedo}}, \citenamefont
  {{Huther}}, \citenamefont {{Langanke}},\ and\ \citenamefont
  {{Heger}}}]{Sieverding.Martinez.ea:2018}%
  \BibitemOpen
  \bibfield  {author} {\bibinfo {author} {\bibfnamefont {A.}~\bibnamefont
  {{Sieverding}}}, \bibinfo {author} {\bibfnamefont {G.}~\bibnamefont
  {{Mart{\'\i}nez-Pinedo}}}, \bibinfo {author} {\bibfnamefont {L.}~\bibnamefont
  {{Huther}}}, \bibinfo {author} {\bibfnamefont {K.}~\bibnamefont
  {{Langanke}}},\ and\ \bibinfo {author} {\bibfnamefont {A.}~\bibnamefont
  {{Heger}}},\ }\bibfield  {title} {\bibinfo {title} {{The
  {\ensuremath{\nu}}-Process in the Light of an Improved Understanding of
  Supernova Neutrino Spectra}},\ }\href
  {https://doi.org/10.3847/1538-4357/aadd48} {\bibfield  {journal} {\bibinfo
  {journal} {Astrophys. J.}\ }\textbf {\bibinfo {volume} {865}},\ \bibinfo
  {eid} {143} (\bibinfo {year} {2018})}\BibitemShut {NoStop}%
\bibitem [{\citenamefont {{Lodders}}()}]{Lodders:2003}%
  \BibitemOpen
  \bibfield  {author} {\bibinfo {author} {\bibfnamefont {K.}~\bibnamefont
  {{Lodders}}},\ }\bibfield  {title} {\bibinfo {title} {{Solar System
  Abundances and Condensation Temperatures of the Elements}},\ }\href@noop {}
  {\bibinfo  {journal} {Astrophys. J.}\ }\BibitemShut {NoStop}%
\bibitem [{\citenamefont {{Woosley}}\ \emph {et~al.}(2002)\citenamefont
  {{Woosley}}, \citenamefont {{Heger}},\ and\ \citenamefont
  {{Weaver}}}]{Woosley.Heger.ea:2002}%
  \BibitemOpen
\bibfield  {journal} {  }\bibfield  {author} {\bibinfo {author} {\bibfnamefont
  {S.~E.}\ \bibnamefont {{Woosley}}}, \bibinfo {author} {\bibfnamefont
  {A.}~\bibnamefont {{Heger}}},\ and\ \bibinfo {author} {\bibfnamefont {T.~A.}\
  \bibnamefont {{Weaver}}},\ }\bibfield  {title} {\bibinfo {title} {{The
  evolution and explosion of massive stars}},\ }\href
  {https://doi.org/10.1103/RevModPhys.74.1015} {\bibfield  {journal} {\bibinfo
  {journal} {Rev. Mod. Phys.}\ }\textbf {\bibinfo {volume} {74}},\ \bibinfo
  {pages} {1015} (\bibinfo {year} {2002})}\BibitemShut {NoStop}%
\bibitem [{\citenamefont {{Weaver}}\ \emph {et~al.}(1978)\citenamefont
  {{Weaver}}, \citenamefont {{Zimmerman}},\ and\ \citenamefont
  {{Woosley}}}]{Weaver.Zimmerman.Woosley:1978}%
  \BibitemOpen
  \bibfield  {author} {\bibinfo {author} {\bibfnamefont {T.~A.}\ \bibnamefont
  {{Weaver}}}, \bibinfo {author} {\bibfnamefont {G.~B.}\ \bibnamefont
  {{Zimmerman}}},\ and\ \bibinfo {author} {\bibfnamefont {S.~E.}\ \bibnamefont
  {{Woosley}}},\ }\bibfield  {title} {\bibinfo {title} {{Presupernova evolution
  of massive stars.}},\ }\href {https://doi.org/10.1086/156569} {\bibfield
  {journal} {\bibinfo  {journal} {Astrophys. J.}\ }\textbf {\bibinfo {volume}
  {225}},\ \bibinfo {pages} {1021} (\bibinfo {year} {1978})}\BibitemShut
  {NoStop}%
\bibitem [{\citenamefont {{Bollig}}\ \emph {et~al.}(2021)\citenamefont
  {{Bollig}}, \citenamefont {{Yadav}}, \citenamefont {{Kresse}}, \citenamefont
  {{Janka}}, \citenamefont {{M{\"u}ller}},\ and\ \citenamefont
  {{Heger}}}]{Bollig.Yadav.ea:2020}%
  \BibitemOpen
  \bibfield  {author} {\bibinfo {author} {\bibfnamefont {R.}~\bibnamefont
  {{Bollig}}}, \bibinfo {author} {\bibfnamefont {N.}~\bibnamefont {{Yadav}}},
  \bibinfo {author} {\bibfnamefont {D.}~\bibnamefont {{Kresse}}}, \bibinfo
  {author} {\bibfnamefont {H.-T.}\ \bibnamefont {{Janka}}}, \bibinfo {author}
  {\bibfnamefont {B.}~\bibnamefont {{M{\"u}ller}}},\ and\ \bibinfo {author}
  {\bibfnamefont {A.}~\bibnamefont {{Heger}}},\ }\bibfield  {title} {\bibinfo
  {title} {{Self-consistent 3D Supernova Models From -7 Minutes to +7 s: A
  1-bethe Explosion of a 19 M$_{{\ensuremath{\odot}}}$ Progenitor}},\ }\href
  {https://doi.org/10.3847/1538-4357/abf82e} {\bibfield  {journal} {\bibinfo
  {journal} {Astrophys. J.}\ }\textbf {\bibinfo {volume} {915}},\ \bibinfo
  {eid} {28} (\bibinfo {year} {2021})}\BibitemShut {NoStop}%
\bibitem [{\citenamefont {{Vartanyan}}\ \emph {et~al.}(2019)\citenamefont
  {{Vartanyan}}, \citenamefont {{Burrows}}, \citenamefont {{Radice}},
  \citenamefont {{Skinner}},\ and\ \citenamefont
  {{Dolence}}}]{Vartanyan.Burrows.ea:2019}%
  \BibitemOpen
  \bibfield  {author} {\bibinfo {author} {\bibfnamefont {D.}~\bibnamefont
  {{Vartanyan}}}, \bibinfo {author} {\bibfnamefont {A.}~\bibnamefont
  {{Burrows}}}, \bibinfo {author} {\bibfnamefont {D.}~\bibnamefont {{Radice}}},
  \bibinfo {author} {\bibfnamefont {M.~A.}\ \bibnamefont {{Skinner}}},\ and\
  \bibinfo {author} {\bibfnamefont {J.}~\bibnamefont {{Dolence}}},\ }\bibfield
  {title} {\bibinfo {title} {{A successful 3D core-collapse supernova explosion
  model}},\ }\href {https://doi.org/10.1093/mnras/sty2585} {\bibfield
  {journal} {\bibinfo  {journal} {Mon. Not. R. Astron. Soc.}\ }\textbf
  {\bibinfo {volume} {482}},\ \bibinfo {pages} {351} (\bibinfo {year}
  {2019})}\BibitemShut {NoStop}%
\bibitem [{\citenamefont {{Lentz}}\ \emph {et~al.}(2015)\citenamefont
  {{Lentz}}, \citenamefont {{Bruenn}}, \citenamefont {{Hix}}, \citenamefont
  {{Mezzacappa}}, \citenamefont {{Messer}}, \citenamefont {{Endeve}},
  \citenamefont {{Blondin}}, \citenamefont {{Harris}}, \citenamefont
  {{Marronetti}},\ and\ \citenamefont {{Yakunin}}}]{Lentz.Bruenn.ea:2015}%
  \BibitemOpen
  \bibfield  {author} {\bibinfo {author} {\bibfnamefont {E.~J.}\ \bibnamefont
  {{Lentz}}}, \bibinfo {author} {\bibfnamefont {S.~W.}\ \bibnamefont
  {{Bruenn}}}, \bibinfo {author} {\bibfnamefont {W.~R.}\ \bibnamefont {{Hix}}},
  \bibinfo {author} {\bibfnamefont {A.}~\bibnamefont {{Mezzacappa}}}, \bibinfo
  {author} {\bibfnamefont {O.~E.~B.}\ \bibnamefont {{Messer}}}, \bibinfo
  {author} {\bibfnamefont {E.}~\bibnamefont {{Endeve}}}, \bibinfo {author}
  {\bibfnamefont {J.~M.}\ \bibnamefont {{Blondin}}}, \bibinfo {author}
  {\bibfnamefont {J.~A.}\ \bibnamefont {{Harris}}}, \bibinfo {author}
  {\bibfnamefont {P.}~\bibnamefont {{Marronetti}}},\ and\ \bibinfo {author}
  {\bibfnamefont {K.~N.}\ \bibnamefont {{Yakunin}}},\ }\bibfield  {title}
  {\bibinfo {title} {{Three-dimensional Core-collapse Supernova Simulated Using
  a 15 M $_{{\ensuremath{\odot}}}$ Progenitor}},\ }\href
  {https://doi.org/10.1088/2041-8205/807/2/L31} {\bibfield  {journal} {\bibinfo
   {journal} {Astrophys. J. Lett.}\ }\textbf {\bibinfo {volume} {807}},\
  \bibinfo {eid} {L31} (\bibinfo {year} {2015})}\BibitemShut {NoStop}%
\bibitem [{\citenamefont {{M{\"u}ller}}\ \emph {et~al.}(2019)\citenamefont
  {{M{\"u}ller}}, \citenamefont {{Tauris}}, \citenamefont {{Heger}},
  \citenamefont {{Banerjee}}, \citenamefont {{Qian}}, \citenamefont {{Powell}},
  \citenamefont {{Chan}}, \citenamefont {{Gay}},\ and\ \citenamefont
  {{Langer}}}]{Mueller.Tauris.ea:2019}%
  \BibitemOpen
  \bibfield  {author} {\bibinfo {author} {\bibfnamefont {B.}~\bibnamefont
  {{M{\"u}ller}}}, \bibinfo {author} {\bibfnamefont {T.~M.}\ \bibnamefont
  {{Tauris}}}, \bibinfo {author} {\bibfnamefont {A.}~\bibnamefont {{Heger}}},
  \bibinfo {author} {\bibfnamefont {P.}~\bibnamefont {{Banerjee}}}, \bibinfo
  {author} {\bibfnamefont {Y.-Z.}\ \bibnamefont {{Qian}}}, \bibinfo {author}
  {\bibfnamefont {J.}~\bibnamefont {{Powell}}}, \bibinfo {author}
  {\bibfnamefont {C.}~\bibnamefont {{Chan}}}, \bibinfo {author} {\bibfnamefont
  {D.~W.}\ \bibnamefont {{Gay}}},\ and\ \bibinfo {author} {\bibfnamefont
  {N.}~\bibnamefont {{Langer}}},\ }\bibfield  {title} {\bibinfo {title}
  {{Three-dimensional simulations of neutrino-driven core-collapse supernovae
  from low-mass single and binary star progenitors}},\ }\href
  {https://doi.org/10.1093/mnras/stz216} {\bibfield  {journal} {\bibinfo
  {journal} {Mon. Not. R. Astron. Soc.}\ }\textbf {\bibinfo {volume} {484}},\
  \bibinfo {pages} {3307} (\bibinfo {year} {2019})}\BibitemShut {NoStop}%
\bibitem [{Note2()}]{Note2}%
  \BibitemOpen
  \bibinfo {note} {\protect \url
  {https://github.com/starkiller-astro/XNet}}\BibitemShut {NoStop}%
\bibitem [{\citenamefont {{Yoshida}}\ \emph {et~al.}(2008)\citenamefont
  {{Yoshida}}, \citenamefont {{Suzuki}}, \citenamefont {{Chiba}}, \citenamefont
  {{Kajino}}, \citenamefont {{Yokomakura}}, \citenamefont {{Kimura}},
  \citenamefont {{Takamura}},\ and\ \citenamefont
  {{Hartmann}}}]{Yoshida.Suzuki.ea:2008}%
  \BibitemOpen
  \bibfield  {author} {\bibinfo {author} {\bibfnamefont {T.}~\bibnamefont
  {{Yoshida}}}, \bibinfo {author} {\bibfnamefont {T.}~\bibnamefont {{Suzuki}}},
  \bibinfo {author} {\bibfnamefont {S.}~\bibnamefont {{Chiba}}}, \bibinfo
  {author} {\bibfnamefont {T.}~\bibnamefont {{Kajino}}}, \bibinfo {author}
  {\bibfnamefont {H.}~\bibnamefont {{Yokomakura}}}, \bibinfo {author}
  {\bibfnamefont {K.}~\bibnamefont {{Kimura}}}, \bibinfo {author}
  {\bibfnamefont {A.}~\bibnamefont {{Takamura}}},\ and\ \bibinfo {author}
  {\bibfnamefont {D.~H.}\ \bibnamefont {{Hartmann}}},\ }\bibfield  {title}
  {\bibinfo {title} {{Neutrino-Nucleus Reaction Cross Sections for Light
  Element Synthesis in Supernova Explosions}},\ }\href
  {https://doi.org/10.1086/591266} {\bibfield  {journal} {\bibinfo  {journal}
  {Astrophys. J.}\ }\textbf {\bibinfo {volume} {686}},\ \bibinfo {pages} {448}
  (\bibinfo {year} {2008})}\BibitemShut {NoStop}%
\bibitem [{\citenamefont {{Kusakabe}}\ \emph {et~al.}(2019)\citenamefont
  {{Kusakabe}}, \citenamefont {{Cheoun}}, \citenamefont {{Kim}}, \citenamefont
  {{Hashimoto}}, \citenamefont {{Ono}}, \citenamefont {{Nomoto}}, \citenamefont
  {{Suzuki}}, \citenamefont {{Kajino}},\ and\ \citenamefont
  {{Mathews}}}]{Kusakabe.Cheoun.ea:2019}%
  \BibitemOpen
  \bibfield  {author} {\bibinfo {author} {\bibfnamefont {M.}~\bibnamefont
  {{Kusakabe}}}, \bibinfo {author} {\bibfnamefont {M.-K.}\ \bibnamefont
  {{Cheoun}}}, \bibinfo {author} {\bibfnamefont {K.~S.}\ \bibnamefont {{Kim}}},
  \bibinfo {author} {\bibfnamefont {M.-a.}\ \bibnamefont {{Hashimoto}}},
  \bibinfo {author} {\bibfnamefont {M.}~\bibnamefont {{Ono}}}, \bibinfo
  {author} {\bibfnamefont {K.}~\bibnamefont {{Nomoto}}}, \bibinfo {author}
  {\bibfnamefont {T.}~\bibnamefont {{Suzuki}}}, \bibinfo {author}
  {\bibfnamefont {T.}~\bibnamefont {{Kajino}}},\ and\ \bibinfo {author}
  {\bibfnamefont {G.~J.}\ \bibnamefont {{Mathews}}},\ }\bibfield  {title}
  {\bibinfo {title} {{Supernova Neutrino Process of Li and B Revisited}},\
  }\href {https://doi.org/10.3847/1538-4357/aafc35} {\bibfield  {journal}
  {\bibinfo  {journal} {Astrophys. J.}\ }\textbf {\bibinfo {volume} {872}},\
  \bibinfo {eid} {164} (\bibinfo {year} {2019})}\BibitemShut {NoStop}%
\bibitem [{\citenamefont {Wu}\ \emph {et~al.}(2015)\citenamefont {Wu},
  \citenamefont {Qian}, \citenamefont {Mart\'{\i}nez-Pinedo}, \citenamefont
  {Fischer},\ and\ \citenamefont {Huther}}]{Wu.Qian.ea:2015}%
  \BibitemOpen
  \bibfield  {author} {\bibinfo {author} {\bibfnamefont {M.-R.}\ \bibnamefont
  {Wu}}, \bibinfo {author} {\bibfnamefont {Y.-Z.}\ \bibnamefont {Qian}},
  \bibinfo {author} {\bibfnamefont {G.}~\bibnamefont {Mart\'{\i}nez-Pinedo}},
  \bibinfo {author} {\bibfnamefont {T.}~\bibnamefont {Fischer}},\ and\ \bibinfo
  {author} {\bibfnamefont {L.}~\bibnamefont {Huther}},\ }\bibfield  {title}
  {\bibinfo {title} {Effects of neutrino oscillations on nucleosynthesis and
  neutrino signals for an $18{M}_{\ensuremath{\bigodot}}$ supernova model},\
  }\href {https://doi.org/10.1103/PhysRevD.91.065016} {\bibfield  {journal}
  {\bibinfo  {journal} {Phys. Rev. C}\ }\textbf {\bibinfo {volume} {91}},\
  \bibinfo {pages} {065016} (\bibinfo {year} {2015})}\BibitemShut {NoStop}%
\bibitem [{\citenamefont {{Yoshida}}\ \emph {et~al.}(2006)\citenamefont
  {{Yoshida}}, \citenamefont {{Kajino}}, \citenamefont {{Yokomakura}},
  \citenamefont {{Kimura}}, \citenamefont {{Takamura}},\ and\ \citenamefont
  {{Hartmann}}}]{Yoshida.Kajino.ea:2006}%
  \BibitemOpen
  \bibfield  {author} {\bibinfo {author} {\bibfnamefont {T.}~\bibnamefont
  {{Yoshida}}}, \bibinfo {author} {\bibfnamefont {T.}~\bibnamefont {{Kajino}}},
  \bibinfo {author} {\bibfnamefont {H.}~\bibnamefont {{Yokomakura}}}, \bibinfo
  {author} {\bibfnamefont {K.}~\bibnamefont {{Kimura}}}, \bibinfo {author}
  {\bibfnamefont {A.}~\bibnamefont {{Takamura}}},\ and\ \bibinfo {author}
  {\bibfnamefont {D.~H.}\ \bibnamefont {{Hartmann}}},\ }\bibfield  {title}
  {\bibinfo {title} {{Supernova Neutrino Nucleosynthesis of Light Elements with
  Neutrino Oscillations}},\ }\href
  {https://doi.org/10.1103/PhysRevLett.96.091101} {\bibfield  {journal}
  {\bibinfo  {journal} {Phys. Rev. Lett.}\ }\textbf {\bibinfo {volume} {96}},\
  \bibinfo {eid} {091101} (\bibinfo {year} {2006})}\BibitemShut {NoStop}%
\bibitem [{\citenamefont {{Sallaska}}\ \emph {et~al.}(2013)\citenamefont
  {{Sallaska}}, \citenamefont {{Iliadis}}, \citenamefont {{Champange}},
  \citenamefont {{Goriely}}, \citenamefont {{Starrfield}},\ and\ \citenamefont
  {{Timmes}}}]{Sallaska.Iliadis.ea:2013}%
  \BibitemOpen
  \bibfield  {author} {\bibinfo {author} {\bibfnamefont {A.~L.}\ \bibnamefont
  {{Sallaska}}}, \bibinfo {author} {\bibfnamefont {C.}~\bibnamefont
  {{Iliadis}}}, \bibinfo {author} {\bibfnamefont {A.~E.}\ \bibnamefont
  {{Champange}}}, \bibinfo {author} {\bibfnamefont {S.}~\bibnamefont
  {{Goriely}}}, \bibinfo {author} {\bibfnamefont {S.}~\bibnamefont
  {{Starrfield}}},\ and\ \bibinfo {author} {\bibfnamefont {F.~X.}\ \bibnamefont
  {{Timmes}}},\ }\bibfield  {title} {\bibinfo {title} {{STARLIB: A
  Next-generation Reaction-rate Library for Nuclear Astrophysics}},\ }\href
  {https://doi.org/10.1088/0067-0049/207/1/18} {\bibfield  {journal} {\bibinfo
  {journal} {Astrophys. J. Suppl. Ser.}\ }\textbf {\bibinfo {volume} {207}},\
  \bibinfo {eid} {18} (\bibinfo {year} {2013})}\BibitemShut {NoStop}%
\bibitem [{\citenamefont {Tilley}\ \emph {et~al.}(2004)\citenamefont {Tilley},
  \citenamefont {Kelley}, \citenamefont {Godwin}, \citenamefont {Millener},
  \citenamefont {Purcell}, \citenamefont {Sheu},\ and\ \citenamefont
  {Weller}}]{Tilley04}%
  \BibitemOpen
  \bibfield  {author} {\bibinfo {author} {\bibfnamefont {D.~R.}\ \bibnamefont
  {Tilley}}, \bibinfo {author} {\bibfnamefont {J.~H.}\ \bibnamefont {Kelley}},
  \bibinfo {author} {\bibfnamefont {J.~L.}\ \bibnamefont {Godwin}}, \bibinfo
  {author} {\bibfnamefont {D.~J.}\ \bibnamefont {Millener}}, \bibinfo {author}
  {\bibfnamefont {J.~E.}\ \bibnamefont {Purcell}}, \bibinfo {author}
  {\bibfnamefont {C.~G.}\ \bibnamefont {Sheu}},\ and\ \bibinfo {author}
  {\bibfnamefont {H.~R.}\ \bibnamefont {Weller}},\ }\bibfield  {title}
  {\bibinfo {title} {{Energy levels of light nuclei $A=8$, 9, 10}},\ }\href
  {https://doi.org/10.1016/j.nuclphysa.2004.09.059} {\bibfield  {journal}
  {\bibinfo  {journal} {Nucl. Phys. A}\ }\textbf {\bibinfo {volume} {745}},\
  \bibinfo {pages} {155} (\bibinfo {year} {2004})}\BibitemShut {NoStop}%
\bibitem [{\citenamefont {Caurier}\ \emph {et~al.}(2005)\citenamefont
  {Caurier}, \citenamefont {Mart\'{\i}nez-Pinedo}, \citenamefont {Nowacki},
  \citenamefont {Poves},\ and\ \citenamefont
  {Zuker}}]{Caurier.Martinez-Pinedo.ea:2005}%
  \BibitemOpen
  \bibfield  {author} {\bibinfo {author} {\bibfnamefont {E.}~\bibnamefont
  {Caurier}}, \bibinfo {author} {\bibfnamefont {G.}~\bibnamefont
  {Mart\'{\i}nez-Pinedo}}, \bibinfo {author} {\bibfnamefont {F.}~\bibnamefont
  {Nowacki}}, \bibinfo {author} {\bibfnamefont {A.}~\bibnamefont {Poves}},\
  and\ \bibinfo {author} {\bibfnamefont {A.~P.}\ \bibnamefont {Zuker}},\
  }\bibfield  {title} {\bibinfo {title} {The shell model as unified view of
  nuclear structure},\ }\href {https://doi.org/10.1103/RevModPhys.77.427}
  {\bibfield  {journal} {\bibinfo  {journal} {Rev. Mod. Phys.}\ }\textbf
  {\bibinfo {volume} {77}},\ \bibinfo {pages} {427} (\bibinfo {year}
  {2005})}\BibitemShut {NoStop}%
\bibitem [{\citenamefont {Cohen}\ and\ \citenamefont {Kurath}(1965)}]{Cohen65}%
  \BibitemOpen
  \bibfield  {author} {\bibinfo {author} {\bibfnamefont {S.}~\bibnamefont
  {Cohen}}\ and\ \bibinfo {author} {\bibfnamefont {D.}~\bibnamefont {Kurath}},\
  }\bibfield  {title} {\bibinfo {title} {Effective interactions for the 1p
  shell},\ }\href {https://doi.org/10.1016/0029-5582(65)90148-3} {\bibfield
  {journal} {\bibinfo  {journal} {Nucl. Phys.}\ }\textbf {\bibinfo {volume}
  {73}},\ \bibinfo {pages} {1} (\bibinfo {year} {1965})}\BibitemShut {NoStop}%
\bibitem [{\citenamefont {Chou}\ \emph {et~al.}(1993)\citenamefont {Chou},
  \citenamefont {Warburton},\ and\ \citenamefont
  {Brown}}]{Chou.Warburton.Brown:1993}%
  \BibitemOpen
  \bibfield  {author} {\bibinfo {author} {\bibfnamefont {W.-T.}\ \bibnamefont
  {Chou}}, \bibinfo {author} {\bibfnamefont {E.~K.}\ \bibnamefont
  {Warburton}},\ and\ \bibinfo {author} {\bibfnamefont {B.~A.}\ \bibnamefont
  {Brown}},\ }\bibfield  {title} {\bibinfo {title} {{Gamow-Teller beta-decay
  rates for $A\leq 18$ nuclei}},\ }\href
  {https://doi.org/10.1103/PhysRevC.47.163} {\bibfield  {journal} {\bibinfo
  {journal} {Phys. Rev. C}\ }\textbf {\bibinfo {volume} {47}},\ \bibinfo
  {pages} {163} (\bibinfo {year} {1993})}\BibitemShut {NoStop}%
\bibitem [{\citenamefont {Ayyad}\ \emph {et~al.}(2019)\citenamefont {Ayyad},
  \citenamefont {Olaizola}, \citenamefont {Mittig}, \citenamefont {Potel},
  \citenamefont {Zelevinsky}, \citenamefont {Horoi}, \citenamefont
  {Beceiro-Novo}, \citenamefont {Alcorta}, \citenamefont {Andreoiu},
  \citenamefont {Ahn}, \citenamefont {Anholm}, \citenamefont {Atar},
  \citenamefont {Babu}, \citenamefont {Bazin}, \citenamefont {Bernier},
  \citenamefont {Bhattacharjee}, \citenamefont {Bowry}, \citenamefont
  {Caballero-Folch}, \citenamefont {Cortesi}, \citenamefont {Dalitz},
  \citenamefont {Dunling}, \citenamefont {Garnsworthy}, \citenamefont {Holl},
  \citenamefont {Kootte}, \citenamefont {Leach}, \citenamefont {Randhawa},
  \citenamefont {Saito}, \citenamefont {Santamaria}, \citenamefont {\ifmmode
  \check{S}\else \v{S}\fi{}iuryt\ifmmode~\dot{e}\else \.{e}\fi{}},
  \citenamefont {Svensson}, \citenamefont {Umashankar}, \citenamefont
  {Watwood},\ and\ \citenamefont {Yates}}]{Ayyad19}%
  \BibitemOpen
  \bibfield  {author} {\bibinfo {author} {\bibfnamefont {Y.}~\bibnamefont
  {Ayyad}}, \bibinfo {author} {\bibfnamefont {B.}~\bibnamefont {Olaizola}},
  \bibinfo {author} {\bibfnamefont {W.}~\bibnamefont {Mittig}}, \bibinfo
  {author} {\bibfnamefont {G.}~\bibnamefont {Potel}}, \bibinfo {author}
  {\bibfnamefont {V.}~\bibnamefont {Zelevinsky}}, \bibinfo {author}
  {\bibfnamefont {M.}~\bibnamefont {Horoi}}, \bibinfo {author} {\bibfnamefont
  {S.}~\bibnamefont {Beceiro-Novo}}, \bibinfo {author} {\bibfnamefont
  {M.}~\bibnamefont {Alcorta}}, \bibinfo {author} {\bibfnamefont
  {C.}~\bibnamefont {Andreoiu}}, \bibinfo {author} {\bibfnamefont
  {T.}~\bibnamefont {Ahn}}, \bibinfo {author} {\bibfnamefont {M.}~\bibnamefont
  {Anholm}}, \bibinfo {author} {\bibfnamefont {L.}~\bibnamefont {Atar}},
  \bibinfo {author} {\bibfnamefont {A.}~\bibnamefont {Babu}}, \bibinfo {author}
  {\bibfnamefont {D.}~\bibnamefont {Bazin}}, \bibinfo {author} {\bibfnamefont
  {N.}~\bibnamefont {Bernier}}, \bibinfo {author} {\bibfnamefont {S.~S.}\
  \bibnamefont {Bhattacharjee}}, \bibinfo {author} {\bibfnamefont
  {M.}~\bibnamefont {Bowry}}, \bibinfo {author} {\bibfnamefont
  {R.}~\bibnamefont {Caballero-Folch}}, \bibinfo {author} {\bibfnamefont
  {M.}~\bibnamefont {Cortesi}}, \bibinfo {author} {\bibfnamefont
  {C.}~\bibnamefont {Dalitz}}, \bibinfo {author} {\bibfnamefont
  {E.}~\bibnamefont {Dunling}}, \bibinfo {author} {\bibfnamefont {A.~B.}\
  \bibnamefont {Garnsworthy}}, \bibinfo {author} {\bibfnamefont
  {M.}~\bibnamefont {Holl}}, \bibinfo {author} {\bibfnamefont {B.}~\bibnamefont
  {Kootte}}, \bibinfo {author} {\bibfnamefont {K.~G.}\ \bibnamefont {Leach}},
  \bibinfo {author} {\bibfnamefont {J.~S.}\ \bibnamefont {Randhawa}}, \bibinfo
  {author} {\bibfnamefont {Y.}~\bibnamefont {Saito}}, \bibinfo {author}
  {\bibfnamefont {C.}~\bibnamefont {Santamaria}}, \bibinfo {author}
  {\bibfnamefont {P.}~\bibnamefont {\ifmmode \check{S}\else
  \v{S}\fi{}iuryt\ifmmode~\dot{e}\else \.{e}\fi{}}}, \bibinfo {author}
  {\bibfnamefont {C.~E.}\ \bibnamefont {Svensson}}, \bibinfo {author}
  {\bibfnamefont {R.}~\bibnamefont {Umashankar}}, \bibinfo {author}
  {\bibfnamefont {N.}~\bibnamefont {Watwood}},\ and\ \bibinfo {author}
  {\bibfnamefont {D.}~\bibnamefont {Yates}},\ }\bibfield  {title} {\bibinfo
  {title} {Direct observation of proton emission in $^{11}\mathrm{Be}$},\
  }\href {https://doi.org/10.1103/PhysRevLett.123.082501} {\bibfield  {journal}
  {\bibinfo  {journal} {Phys. Rev. Lett.}\ }\textbf {\bibinfo {volume} {123}},\
  \bibinfo {pages} {082501} (\bibinfo {year} {2019})}\BibitemShut {NoStop}%
\bibitem [{\citenamefont {Oko\l{}owicz}\ \emph {et~al.}(2020)\citenamefont
  {Oko\l{}owicz}, \citenamefont {P\l{}oszajczak},\ and\ \citenamefont
  {Nazarewicz}}]{Okol20}%
  \BibitemOpen
  \bibfield  {author} {\bibinfo {author} {\bibfnamefont {J.}~\bibnamefont
  {Oko\l{}owicz}}, \bibinfo {author} {\bibfnamefont {M.}~\bibnamefont
  {P\l{}oszajczak}},\ and\ \bibinfo {author} {\bibfnamefont {W.}~\bibnamefont
  {Nazarewicz}},\ }\bibfield  {title} {\bibinfo {title} {Convenient location of
  a near-threshold proton-emitting resonance in $^{11}\mathrm{B}$},\ }\href
  {https://doi.org/10.1103/PhysRevLett.124.042502} {\bibfield  {journal}
  {\bibinfo  {journal} {Phys. Rev. Lett.}\ }\textbf {\bibinfo {volume} {124}},\
  \bibinfo {pages} {042502} (\bibinfo {year} {2020})}\BibitemShut {NoStop}%
\bibitem [{\citenamefont {Azuma}\ \emph {et~al.}(2010)\citenamefont {Azuma},
  \citenamefont {Uberseder}, \citenamefont {Simpson}, \citenamefont {Brune},
  \citenamefont {Costantini}, \citenamefont {de~Boer}, \citenamefont
  {G\"orres}, \citenamefont {Heil}, \citenamefont {LeBlanc}, \citenamefont
  {Ugalde},\ and\ \citenamefont {Wiescher}}]{azure}%
  \BibitemOpen
  \bibfield  {author} {\bibinfo {author} {\bibfnamefont {R.~E.}\ \bibnamefont
  {Azuma}}, \bibinfo {author} {\bibfnamefont {E.}~\bibnamefont {Uberseder}},
  \bibinfo {author} {\bibfnamefont {E.~C.}\ \bibnamefont {Simpson}}, \bibinfo
  {author} {\bibfnamefont {C.~R.}\ \bibnamefont {Brune}}, \bibinfo {author}
  {\bibfnamefont {H.}~\bibnamefont {Costantini}}, \bibinfo {author}
  {\bibfnamefont {R.~J.}\ \bibnamefont {de~Boer}}, \bibinfo {author}
  {\bibfnamefont {J.}~\bibnamefont {G\"orres}}, \bibinfo {author}
  {\bibfnamefont {M.}~\bibnamefont {Heil}}, \bibinfo {author} {\bibfnamefont
  {P.~J.}\ \bibnamefont {LeBlanc}}, \bibinfo {author} {\bibfnamefont
  {C.}~\bibnamefont {Ugalde}},\ and\ \bibinfo {author} {\bibfnamefont
  {M.}~\bibnamefont {Wiescher}},\ }\bibfield  {title} {\bibinfo {title}
  {{AZURE: An $R$-matrix code for nuclear astrophysics}},\ }\href
  {https://doi.org/10.1103/PhysRevC.81.045805} {\bibfield  {journal} {\bibinfo
  {journal} {Phys. Rev. C}\ }\textbf {\bibinfo {volume} {81}},\ \bibinfo
  {pages} {045805} (\bibinfo {year} {2010})}\BibitemShut {NoStop}%
\bibitem [{\citenamefont {Uberseder}\ and\ \citenamefont
  {deBoer}(2015)}]{azure2}%
  \BibitemOpen
  \bibfield  {author} {\bibinfo {author} {\bibfnamefont {E.}~\bibnamefont
  {Uberseder}}\ and\ \bibinfo {author} {\bibfnamefont {R.~J.}\ \bibnamefont
  {deBoer}},\ }\href {azure.nd.edu} {\emph {\bibinfo {title} {{\texttt{AZURE2}
  User Manual}}}} (\bibinfo {year} {2015}),\ \bibinfo {note}
  {azure.nd.edu}\BibitemShut {NoStop}%
\bibitem [{\citenamefont {Refsgaard}\ \emph {et~al.}(2019)\citenamefont
  {Refsgaard}, \citenamefont {B\"uscher}, \citenamefont {Arokiaraj},
  \citenamefont {Fynbo}, \citenamefont {Raabe},\ and\ \citenamefont
  {Riisager}}]{Jonas19}%
  \BibitemOpen
  \bibfield  {author} {\bibinfo {author} {\bibfnamefont {J.}~\bibnamefont
  {Refsgaard}}, \bibinfo {author} {\bibfnamefont {J.}~\bibnamefont
  {B\"uscher}}, \bibinfo {author} {\bibfnamefont {A.}~\bibnamefont
  {Arokiaraj}}, \bibinfo {author} {\bibfnamefont {H.~O.~U.}\ \bibnamefont
  {Fynbo}}, \bibinfo {author} {\bibfnamefont {R.}~\bibnamefont {Raabe}},\ and\
  \bibinfo {author} {\bibfnamefont {K.}~\bibnamefont {Riisager}},\ }\bibfield
  {title} {\bibinfo {title} {Clarification of large-strength transitions in the
  $\ensuremath{\beta}$ decay of $^{11}\mathrm{Be}$},\ }\href
  {https://doi.org/10.1103/PhysRevC.99.044316} {\bibfield  {journal} {\bibinfo
  {journal} {Phys. Rev. C}\ }\textbf {\bibinfo {volume} {99}},\ \bibinfo
  {pages} {044316} (\bibinfo {year} {2019})}\BibitemShut {NoStop}%
\bibitem [{NND()}]{NNDC}%
  \BibitemOpen
  \href@noop {} {}\bibinfo {note}
  {\url{https://www.nndc.bnl.gov/nudat3/getdataset.jsp?nucleus=11B&unc=NDS}}\BibitemShut
  {NoStop}%
\bibitem [{\citenamefont {Angulo}\ \emph {et~al.}(1999)\citenamefont {Angulo},
  \citenamefont {Arnould}, \citenamefont {Rayet}, \citenamefont {Descouvemont},
  \citenamefont {Baye}, \citenamefont {Leclercq-Willain}, \citenamefont {Coc},
  \citenamefont {Barhoumi}, \citenamefont {Aguer}, \citenamefont {Rolfs},
  \citenamefont {Kunz}, \citenamefont {Hammer}, \citenamefont {Mayer},
  \citenamefont {Paradellis}, \citenamefont {Kossionides}, \citenamefont
  {Chronidou}, \citenamefont {Spyrou}, \citenamefont {Degl'Innocenti},
  \citenamefont {Fiorentini}, \citenamefont {Ricci}, \citenamefont
  {Zavatarelli}, \citenamefont {Providencia}, \citenamefont {Wolters},
  \citenamefont {Soares}, \citenamefont {Grama}, \citenamefont {Rahighi},
  \citenamefont {Shotter},\ and\ \citenamefont {{Lamehi
  Rachti}}}]{ANGULO19993}%
  \BibitemOpen
  \bibfield  {author} {\bibinfo {author} {\bibfnamefont {C.}~\bibnamefont
  {Angulo}}, \bibinfo {author} {\bibfnamefont {M.}~\bibnamefont {Arnould}},
  \bibinfo {author} {\bibfnamefont {M.}~\bibnamefont {Rayet}}, \bibinfo
  {author} {\bibfnamefont {P.}~\bibnamefont {Descouvemont}}, \bibinfo {author}
  {\bibfnamefont {D.}~\bibnamefont {Baye}}, \bibinfo {author} {\bibfnamefont
  {C.}~\bibnamefont {Leclercq-Willain}}, \bibinfo {author} {\bibfnamefont
  {A.}~\bibnamefont {Coc}}, \bibinfo {author} {\bibfnamefont {S.}~\bibnamefont
  {Barhoumi}}, \bibinfo {author} {\bibfnamefont {P.}~\bibnamefont {Aguer}},
  \bibinfo {author} {\bibfnamefont {C.}~\bibnamefont {Rolfs}}, \bibinfo
  {author} {\bibfnamefont {R.}~\bibnamefont {Kunz}}, \bibinfo {author}
  {\bibfnamefont {J.}~\bibnamefont {Hammer}}, \bibinfo {author} {\bibfnamefont
  {A.}~\bibnamefont {Mayer}}, \bibinfo {author} {\bibfnamefont
  {T.}~\bibnamefont {Paradellis}}, \bibinfo {author} {\bibfnamefont
  {S.}~\bibnamefont {Kossionides}}, \bibinfo {author} {\bibfnamefont
  {C.}~\bibnamefont {Chronidou}}, \bibinfo {author} {\bibfnamefont
  {K.}~\bibnamefont {Spyrou}}, \bibinfo {author} {\bibfnamefont
  {S.}~\bibnamefont {Degl'Innocenti}}, \bibinfo {author} {\bibfnamefont
  {G.}~\bibnamefont {Fiorentini}}, \bibinfo {author} {\bibfnamefont
  {B.}~\bibnamefont {Ricci}}, \bibinfo {author} {\bibfnamefont
  {S.}~\bibnamefont {Zavatarelli}}, \bibinfo {author} {\bibfnamefont
  {C.}~\bibnamefont {Providencia}}, \bibinfo {author} {\bibfnamefont
  {H.}~\bibnamefont {Wolters}}, \bibinfo {author} {\bibfnamefont
  {J.}~\bibnamefont {Soares}}, \bibinfo {author} {\bibfnamefont
  {C.}~\bibnamefont {Grama}}, \bibinfo {author} {\bibfnamefont
  {J.}~\bibnamefont {Rahighi}}, \bibinfo {author} {\bibfnamefont
  {A.}~\bibnamefont {Shotter}},\ and\ \bibinfo {author} {\bibfnamefont
  {M.}~\bibnamefont {{Lamehi Rachti}}},\ }\bibfield  {title} {\bibinfo {title}
  {A compilation of charged-particle induced thermonuclear reaction rates},\
  }\href {https://doi.org/https://doi.org/10.1016/S0375-9474(99)00030-5}
  {\bibfield  {journal} {\bibinfo  {journal} {Nucl. Phys. A}\ }\textbf
  {\bibinfo {volume} {656}},\ \bibinfo {pages} {3} (\bibinfo {year}
  {1999})}\BibitemShut {NoStop}%
\bibitem [{\citenamefont {Mittig}\ and\ \citenamefont
  {Ayyad}(2022)}]{Mittig_Yassid}%
  \BibitemOpen
  \bibfield  {author} {\bibinfo {author} {\bibfnamefont {W.}~\bibnamefont
  {Mittig}}\ and\ \bibinfo {author} {\bibfnamefont {Y.}~\bibnamefont {Ayyad}},\
  }\href@noop {} {}\bibinfo {howpublished} {In prep.} (\bibinfo {year}
  {2022})\BibitemShut {NoStop}%
\bibitem [{\citenamefont {{Wasserburg}}\ \emph {et~al.}(2006)\citenamefont
  {{Wasserburg}}, \citenamefont {{Busso}}, \citenamefont {{Gallino}},\ and\
  \citenamefont {{Nollett}}}]{Wasserburg.Busso.ea:2006}%
  \BibitemOpen
  \bibfield  {author} {\bibinfo {author} {\bibfnamefont {G.~J.}\ \bibnamefont
  {{Wasserburg}}}, \bibinfo {author} {\bibfnamefont {M.}~\bibnamefont
  {{Busso}}}, \bibinfo {author} {\bibfnamefont {R.}~\bibnamefont {{Gallino}}},\
  and\ \bibinfo {author} {\bibfnamefont {K.~M.}\ \bibnamefont {{Nollett}}},\
  }\bibfield  {title} {\bibinfo {title} {{Short-lived nuclei in the early Solar
  System: Possible AGB sources}},\ }\href
  {https://doi.org/10.1016/j.nuclphysa.2005.07.015} {\bibfield  {journal}
  {\bibinfo  {journal} {Nucl. Phys. A}\ }\textbf {\bibinfo {volume} {777}},\
  \bibinfo {pages} {5} (\bibinfo {year} {2006})}\BibitemShut {NoStop}%
\bibitem [{\citenamefont {{Asplund}}\ \emph {et~al.}(2009)\citenamefont
  {{Asplund}}, \citenamefont {{Grevesse}}, \citenamefont {{Sauval}},\ and\
  \citenamefont {{Scott}}}]{Asplund.Grevesse.ea:2009}%
  \BibitemOpen
  \bibfield  {author} {\bibinfo {author} {\bibfnamefont {M.}~\bibnamefont
  {{Asplund}}}, \bibinfo {author} {\bibfnamefont {N.}~\bibnamefont
  {{Grevesse}}}, \bibinfo {author} {\bibfnamefont {A.~J.}\ \bibnamefont
  {{Sauval}}},\ and\ \bibinfo {author} {\bibfnamefont {P.}~\bibnamefont
  {{Scott}}},\ }\bibfield  {title} {\bibinfo {title} {{The Chemical Composition
  of the Sun}},\ }\href
  {https://doi.org/10.1146/annurev.astro.46.060407.145222} {\bibfield
  {journal} {\bibinfo  {journal} {Annu. Rev. Astron. Astrophys.}\ }\textbf
  {\bibinfo {volume} {47}},\ \bibinfo {pages} {481} (\bibinfo {year}
  {2009})}\BibitemShut {NoStop}%
\bibitem [{\citenamefont {{Liu}}\ \emph {et~al.}(2012)\citenamefont {{Liu}},
  \citenamefont {{Chaussidon}}, \citenamefont {{Srinivasan}},\ and\
  \citenamefont {{McKeegan}}}]{Liu.Chaussidon.ea:2012}%
  \BibitemOpen
  \bibfield  {author} {\bibinfo {author} {\bibfnamefont {M.-C.}\ \bibnamefont
  {{Liu}}}, \bibinfo {author} {\bibfnamefont {M.}~\bibnamefont {{Chaussidon}}},
  \bibinfo {author} {\bibfnamefont {G.}~\bibnamefont {{Srinivasan}}},\ and\
  \bibinfo {author} {\bibfnamefont {K.~D.}\ \bibnamefont {{McKeegan}}},\
  }\bibfield  {title} {\bibinfo {title} {{A Lower Initial Abundance of
  Short-lived $^{41}$Ca in the Early Solar System and Its Implications for
  Solar System Formation}},\ }\href
  {https://doi.org/10.1088/0004-637X/761/2/137} {\bibfield  {journal} {\bibinfo
   {journal} {Astrophys. J.}\ }\textbf {\bibinfo {volume} {761}},\ \bibinfo
  {eid} {137} (\bibinfo {year} {2012})}\BibitemShut {NoStop}%
\bibitem [{\citenamefont {van~der Walt}\ \emph {et~al.}(2011)\citenamefont
  {van~der Walt}, \citenamefont {Colbert},\ and\ \citenamefont
  {Varoquaux}}]{numpy1}%
  \BibitemOpen
  \bibfield  {author} {\bibinfo {author} {\bibfnamefont {S.}~\bibnamefont
  {van~der Walt}}, \bibinfo {author} {\bibfnamefont {S.~C.}\ \bibnamefont
  {Colbert}},\ and\ \bibinfo {author} {\bibfnamefont {G.}~\bibnamefont
  {Varoquaux}},\ }\bibfield  {title} {\bibinfo {title} {The numpy array: A
  structure for efficient numerical computation},\ }\href
  {https://doi.org/10.1109/MCSE.2011.37} {\bibfield  {journal} {\bibinfo
  {journal} {Computing in Science Engineering}\ }\textbf {\bibinfo {volume}
  {13}},\ \bibinfo {pages} {22} (\bibinfo {year} {2011})}\BibitemShut {NoStop}%
\bibitem [{\citenamefont {{Hunter}}(2007)}]{Hunter:2007}%
  \BibitemOpen
  \bibfield  {author} {\bibinfo {author} {\bibfnamefont {J.~D.}\ \bibnamefont
  {{Hunter}}},\ }\bibfield  {title} {\bibinfo {title} {{Matplotlib: A 2D
  Graphics Environment}},\ }\href {https://doi.org/10.1109/MCSE.2007.55}
  {\bibfield  {journal} {\bibinfo  {journal} {Computing in Science and
  Engineering}\ }\textbf {\bibinfo {volume} {9}},\ \bibinfo {pages} {90}
  (\bibinfo {year} {2007})}\BibitemShut {NoStop}%
\bibitem [{\citenamefont {{Rauscher}}\ and\ \citenamefont
  {{Thielemann}}(2000)}]{Rauscher.Thielemann:2000}%
  \BibitemOpen
  \bibfield  {author} {\bibinfo {author} {\bibfnamefont {T.}~\bibnamefont
  {{Rauscher}}}\ and\ \bibinfo {author} {\bibfnamefont {F.-K.}\ \bibnamefont
  {{Thielemann}}},\ }\bibfield  {title} {\bibinfo {title} {{Astrophysical
  Reaction Rates From Statistical Model Calculations}},\ }\href
  {https://doi.org/10.1006/adnd.2000.0834} {\bibfield  {journal} {\bibinfo
  {journal} {At. Data Nucl. Data Tables}\ }\textbf {\bibinfo {volume} {75}},\
  \bibinfo {pages} {1} (\bibinfo {year} {2000})}\BibitemShut {NoStop}%
\end{thebibliography}%

\end{document}